\documentclass[usegraphicx,usenatbib]{mn2e}
\usepackage{afterpage}
\title[The dynamics of isolated Local Group galaxies]{The dynamics of
  isolated Local Group galaxies\thanks{The data presented herein were
    obtained at the W.~M.~Keck Observatory, which is operated as a
    scientific partnership among the California Institute of
    Technology, the University of California and the National
    Aeronautics and Space Administration. The Observatory was made
    possible by the generous financial support of the W.~M.~Keck
    Foundation.}}

\author[E. N. Kirby et al.]{Evan~N.~Kirby,$^{1}$\thanks{Center for Galaxy Evolution Fellow.}\thanks{E-mail: ekirby@uci.edu}
  James~S.~Bullock,$^{1}$ Michael Boylan-Kolchin,$^{2}$ \newauthor Manoj~Kaplinghat$^{1}$ and Judith~G.~Cohen$^{3}$ \\
$^{1}$Department of Physics and Astronomy, University of California, 4129 Frederick Reines Hall, Irvine, California 92697, USA \\
$^{2}$Department of Astronomy and Joint Space-Science Institute, University of Maryland, College Park, Maryland 20742, USA \\
$^{3}$Department of Astronomy \& Astrophysics, California Institute of Technology, 1200 E.\ California Blvd., MC 249-17, Pasadena, \\ CA 91125, USA}
\begin{document}

\newcommand{\icosotvmean}{-231.6}
\newcommand{\icosotsigmav}{10.8}
\newcommand{\icosotvlow}{-259.5}
\newcommand{\icosotvhigh}{-203.8}
\newcommand{\icosotstarv}{-198.0}
\newcommand{\icosotstarverr}{14.1}
\newcommand{\nsettsigmav}{ 23.2}
\newcommand{\nsettsigmaverr}{1.2}
\newcommand{\leoasigmav}{6.7}
\newcommand{\leoasigmaverru}{1.4}
\newcommand{\leoasigmaverrl}{1.2}
\newcommand{\leoasigmavsigmadiff}{1}
\newcommand{\cetsigmav}{8.3}
\newcommand{\cetsigmaverru}{1.0}
\newcommand{\cetsigmaverrl}{1.0}
\newcommand{\cetsigmavsigmadiff}{4}
\newcommand{\brosigmav}{10.8}
\newcommand{\brosigmaverru}{3.7}
\newcommand{\brosigmaverrl}{2.8}
\newcommand{\brosigmavsigmadiff}{1.3}
\newcommand{\lewsigmav}{12.0}
\newcommand{\lewsigmaverru}{2.0}
\newcommand{\lewsigmaverrl}{1.9}
\newcommand{\lewsigmavsigmadiff}{1.7}
\newcommand{\nlew}{70}
\newcommand{\nlewmatch}{23}
\newcommand{\nlewmatchoff}{*}
\newcommand{\nlewmatchon}{20}
\newcommand{\ksallrh}{51}
\newcommand{\ksallrhmod}{65}
\newcommand{\ksalllogrh}{ 2}
\newcommand{\ksalllogrhmod}{12}
\newcommand{\ksalllum}{95}
\newcommand{\ksalllummod}{89}
\newcommand{\nstars}{862}

\newcommand{\pegvrot}{10.0}
\newcommand{\pegvroterr}{0.3}

\newcommand{\msun}{{\rmn M}_{\sun}}
\newcommand{\lsun}{{\rmn L}_{\sun}}

\date{Accepted 2014 January 6.  Received 2013 December 20; in original form 2013 October 31}

\pagerange{\pageref{firstpage}--\pageref{lastpage}} \pubyear{2014}

\maketitle

\label{firstpage}


\begin{abstract}
We measured velocities of \nstars\ individual red giant stars in seven
isolated dwarf galaxies in the Local Group: NGC~6822, IC~1613, VV~124
(UGC~4879), the Pegasus dwarf irregular galaxy (DDO~216), Leo~A,
Cetus, and Aquarius (DDO~210).  We also computed velocity dispersions,
taking into account the measurement uncertainties on individual stars.
None of the isolated galaxies is denser than the densest Local Group
satellite galaxy.  Furthermore, the isolated dwarf galaxies have no
obvious distinction in the velocity dispersion--half-light radius
plane from the satellite galaxies of the Milky Way and M31.  The
similarity of the isolated and satellite galaxies' dynamics and
structural parameters imposes limitations on environmental solutions
to the too-big-to-fail problem, wherein there are fewer dense dwarf
satellite galaxies than would be expected from cold dark matter
simulations.  This data set also has many other applications for dwarf
galaxy evolution, including the transformation of dwarf irregular into
dwarf spheroidal galaxies.  We intend to explore these issues in
future work.
\end{abstract}

\begin{keywords}
Local Group -- galaxies: dwarf -- galaxies: kinematics and dynamics.
\end{keywords}


\section{Introduction}
\label{sec:intro}

Nearby dwarf galaxies are excellent laboratories to study galactic
dynamics, chemical evolution and dark matter physics.  They lend
themselves so well to detailed scrutiny because they are close enough
for resolved stellar spectroscopy.  Multi-object spectrographs on
8--10~m telescopes, like Keck/DEIMOS \citep{fab03}, Gemini/GMOS
\citep{hoo04}, VLT/FORS2 \citep{app98} and VLT/FLAMES \citep{pas02},
can amass sample sizes of hundreds of stellar spectra for galaxies as
far away as $\sim 1.5$~Mpc.  Even with low signal-to-noise (S/N)
ratios, it is possible to measure radial velocities with typical
precisions of 2--5~km~s$^{-1}$.  Higher quality spectra permit
measurements of metallicities \citep[e.g.,][]{tol01} and even detailed
abundance ratios \citep[e.g.,][]{kir09,let10}.

The Milky Way's satellite galaxies are the best-studied dwarfs in the
Local Volume because of their proximity.  Recent large surveys like
the Spectroscopic and Panchromatic Landscape of Andromeda's Stellar
Halo \citep[SPLASH,][]{guh05,guh06}, the Pan-Andromeda Archaeological
Survey \citep[PAndAS,][]{mcc09} and the Panchromatic Hubble Andromeda
Treasury \citep[PHAT,][]{dal12} have increased the accessibility of
M31 and its satellites.  The isolated dwarf galaxies in the field of
the Local Group are more difficult to discover and to observe because
they are more distant than the Milky Way satellites, and they span the
entire sky, unlike the M31 satellites.

The structural properties of dwarf galaxies may be influenced by their
environments.  For example, a large host galaxy may induce tidal
stripping and deformation \citep*[e.g.,][]{pen08}.  Proximity to a
large galaxy also strongly correlates with a dwarf galaxy's gas
content \citep*[e.g.,][]{kna78,grc09}, star formation history
\citep[e.g.,][]{gre03,wei11} and morphology
\citep*[e.g.,][]{bin90,lis07}.  The isolation of Local Group galaxies
that are far from the Milky Way or M31 insulates them from tides and
ram pressure.  It is possible that some of the isolated Local Group
dwarf galaxies passed near a large galaxy but did not become bound
\citep*[`backsplash galaxies,'][]{sal07,tey12}.  An otherwise isolated
dwarf galaxy can also interact with another dwarf galaxy or even a
dark subhalo \citep{hel12}.  Despite these possible disturbances, the
field of the Local Group is the best place to study the evolution of
dwarf galaxies that have survived unmolested for the age of the
Universe.

There are compelling reasons for comparing the dynamical properties of
satellite galaxies to isolated galaxies.  Such comparisons are likely
to shed light on formation and evolution mechanisms for both dwarf
irregular (dIrr) and dwarf spheroidal (dSph) galaxies.  It is now
known that these two classes of galaxies share the same stellar
mass--stellar metallicity relation \citep{kir13b}, which limits the
amount of stellar stripping associated with possible transformations
of dIrrs to dSphs.  Dynamical measurements should provide further and
complementary constraints.

Furthermore, comparing the dynamics of isolated galaxies to satellite
galaxies can inform possible solutions to the `too big to fail'
problem \citep*[TBTF,][]{boy11,boy12}.  Cold dark matter simulations
predict more dense dark matter subhaloes than are observed among the
Milky Way satellite galaxies.  The problem also exists for the M31
system \citep{tol12,tol13,col13}.  One way to alleviate TBTF is to
invoke baryonic physics, wherein energy injection from gas and stars
alters the mass profiles of the dark matter subhaloes
\citep*[e.g.,][]{bro13}.  High-resolution hydrodynamical simulations
\citep[e.g.,][]{zol12}, analytic arguments \citep{pen12} and idealized
numerical simulations \citep{gar13} indicate energy injection alone is
unlikely to fully explain TBTF because the low stellar content of the
Milky Way satellites places a strong limit on the amount of feedback
available \citep*[though see][]{amo13}.  Environmental effects such as
tides and ram pressure are therefore central to baryonic solutions to
TBTF \citep{arr13,brozol12}.  For that reason, it is important to
measure the kinematic and structural properties for isolated dwarf
galaxies in the field of the Local Group and compare them to dwarf
satellites.

Stellar velocity measurements are already available for seven isolated
dwarf galaxies.  \citet{tol01} measured the velocities for 23~red
giants in the dIrr NGC~6822, but the primary purpose of their survey
was to measure the metallicity distribution.  \citet*{dem06} also
measured the velocities of 110~carbon stars in NGC~6822, but the
intrinsic variability of the stars limited the velocity precision to
$\sim 15$~km~s$^{-1}$, larger than the velocity dispersion of any
Milky Way dSph.  \citet{lea09,lea12} measured the velocities of
180~red giants in the WLM dIrr.  They found a stellar rotation
velocity about equal to the velocity dispersion.  They also calculated
a very large mass within the half-light radius ($(4.3 \pm 0.3) \times
10^8~\msun$), which is perhaps appropriate for its large half-light
radius (1.6~kpc).  \citet*{kir12} observed 67~red giants in VV~124
(UGC~4879).  They found that most of the mass within the half-light
radius of VV~124 is dark matter despite the relatively small
half-light radius (272~pc).  \citet{bro07} observed ten B supergiants
in Leo~A\@.  From these young stars, they determined that Leo~A's
dynamical mass is at least five times its stellar mass.  \citet{lew07}
also measured a very high dynamical mass-to-light ratio for the Cetus
dSph.  However, the quality of their spectra yielded a mean velocity
uncertainty of 8.8~km~s$^{-1}$.  \citet{fra09} measured both
dispersion and rotation in Tucana.  Both values are similar to those
for WLM, which is surprising given Tucana's low luminosity ($6 \times
10^5~\lsun$).  Finally, \citet{sim07} measured the velocity dispersion
for Leo~T and seven other ultra-faint dwarf galaxies.  Leo~T is
probably the only recently discovered \citep{irw07} ultra-faint galaxy
near the Milky Way that is on its first infall.  Regardless, its
mass-to-light ratio is very high ($\sim 90~\msun~\lsun^{-1}$), like
the other ultra-faint galaxies.

In this contribution, we expand the kinematic observations of isolated
dwarf galaxies.  We refine the kinematic measurements for NGC~6822,
Leo~A and Cetus.  We also provide velocity dispersions for isolated
dwarf galaxies without previous measurements: IC~1613, the Pegasus
dIrr (DDO~216) and Aquarius (DDO~210).  We also include our previous
kinematic measurements of VV~124 \citep{kir12}.  In
Secs.~\ref{sec:obs} and \ref{sec:vel}, we describe our spectroscopy
and measurements of individual stellar velocities.  We provide a table
of velocities so that others may construct their own dynamical models
of these galaxies.  We rule stars as members and non-members of their
respective galaxies in Sec.~\ref{sec:membership}.  In
Sec.~\ref{sec:disp}, we describe the calculation of velocity
dispersions.  Finally, we compare the kinematic and structural
properties of isolated dwarf galaxies to dwarf satellites of the Milky
Way and M31 in Sec.~\ref{sec:discussion}.


\section{Observations}
\label{sec:obs}

\begin{table*}
\centering
\begin{minipage}{110mm}
\caption{Summary of DEIMOS observations.}
\label{tab:obs}
\begin{tabular}{llrrl}
\hline
Galaxy & Slitmask & \# Targets & Exposure Time (h) & Originally Published By \\
\hline
IC~1613      & i1613a  &    199 &    10.3 & \citet{kir13b} \\
NGC~6822     & n6822a  &    180 &     8.7 & \citet{kir13b} \\
             & n6822b  &    180 &     6.0 & \citet{kir13b} \\
VV~124       & vv124a  &    121 &     3.7 & \citet{kir12} \\
             & vv124b  &    120 &     3.8 & \citet{kir12} \\
Pegasus      & pega    &    113 &     6.8 & \citet{kir13b} \\
Leo~A        & leoaaW  &     91 &     6.7 & \citet{kir13b} \\
Cetus        & ceta    &    146 &     5.8 & this work \\
             & cetb    &    131 &     5.8 & this work \\
Aquarius     & aqra    &     64 &     8.9 & \citet{kir13b} \\
\hline
\end{tabular}
\end{minipage}
\end{table*}

We observed seven isolated galaxies in the Local Group with the Deep
Imaging Multi-object Spectrograph \citep[DEIMOS,][]{fab03} on the
Keck~II telescope.  Table~\ref{tab:obs} summarises our observations.
\citet{kir12,kir13b} already presented the observations of most these
galaxies.  We add observations of Cetus to these published data.

\begin{figure}
 \centering
 \includegraphics[width=85mm]{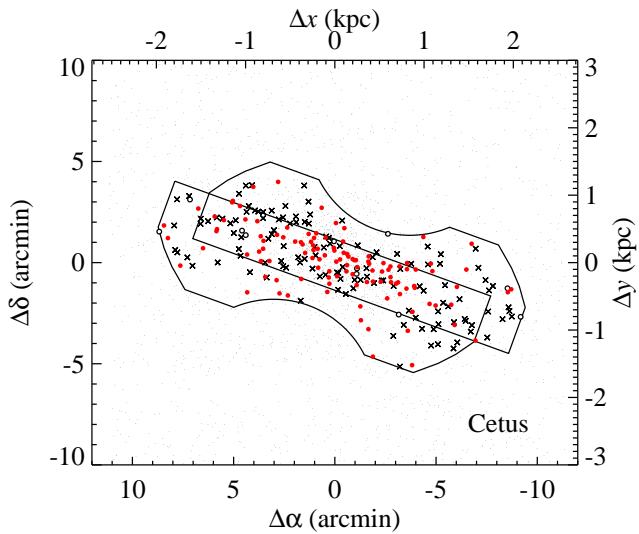}
 \caption{The sky position of the two Keck/DEIMOS slitmasks for Cetus.
   Filled, red points are spectroscopically confirmed members.
   Hollow, black points are non-members.  Crosses are spectroscopic
   targets for which it was not possible to measure a velocity.  The
   origin is $\alpha_0 = 00^{\rmn h} 26^{\rmn m} 11^{\rmn s}$,
   $\delta_0 = -11\degr 02\arcmin 40\arcsec$.  The top and right axes
   give the projected distance from the centre of Cetus in kpc for an
   assumed distance of 779~kpc
   \protect\citep{ber09}.\label{fig:cetcoords}}
\end{figure}

We selected Cetus targets in the same manner as the other dwarf
galaxies \citep{kir12,kir13b}.  A.~McConnachie and M.~Irwin kindly
provided us their photometric catalogue from the Wide Field Camera on
the Isaac Newton Telescope \citep{mcc06}.  We designed two DEIMOS
slitmasks centred on Cetus.  Fig.~\ref{fig:cetcoords} shows the
positions and orientations of the slitmasks in celestial coordinates.
It also identifies stars that we later determined to be spectroscopic
members and non-members (Sec.~\ref{sec:membership}).

\begin{figure}
 \centering
 \includegraphics[width=85mm]{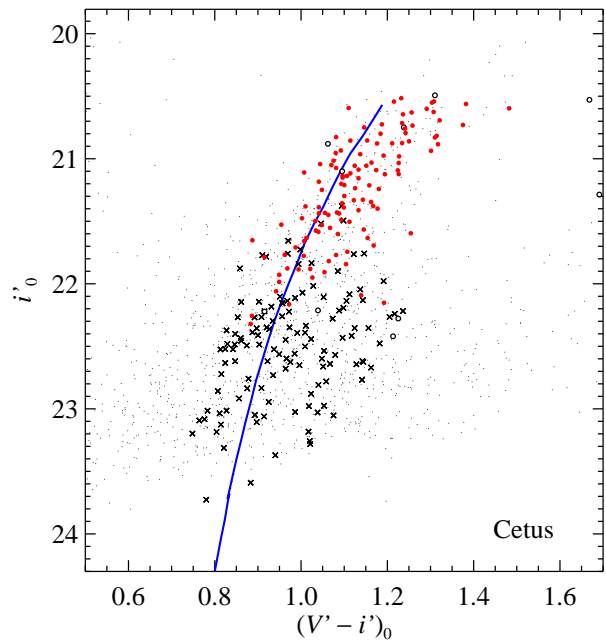}
 \caption{The extinction- and reddening-corrected colour--magnitude
   diagram for Cetus from the photometry catalogue of
   \protect\citet{mcc06}.  The filters are Johnson $V'$ and Gunn $i'$.
   Symbols have the same meanings as Fig.~\ref{fig:cetcoords}.  The
   blue line is a 12.6~Gyr Padova isochrone \protect\citep{gir02}
   with $\rmn{[Fe/H]} = -1.6$ at a distance modulus of $(m-M)_0 =
   24.46$ \protect\citep{ber09}.\label{fig:cetcmd}}
\end{figure}

Targets were selected to have the colours and magnitudes of red
giants, assuming a distance modulus of $(m-M)_0 = 24.46$ \citep{ber09}
and a reddening of $E(B-V) = 0.029$ \citep*{sch98}.  In practice, we
chose stars with $20.4 < i'_0 < 24.0$ and colours within 0.3~mag of a
12.6~Gyr Padova isochrone \citep{gir02} with $\rmn{[Fe/H]} = -1.6$.
We then added other objects outside of this selection region in order
to fill the slitmask.  Fig.~\ref{fig:cetcmd} shows the
colour--magnitude diagram from the photometric catalogue
\citep{mcc06}.  Spectroscopic targets, including members and
non-members (determined as described in Sec.~\ref{sec:membership}) are
indicated with large symbols.

We observed the first Cetus slitmask, ceta, on 2013 September 1.  We
observed the second slitmask, cetb, on the next night.  We obtained 12
exposures for each slitmask with a total exposure time of 350~minutes
(5.8~hours) per slitmask.  The sky was clear on both nights.  The
seeing was about $0\farcs 9$ for the first night, but the seeing
varied between $1\farcs 0$ and $1\farcs 5$ for most of the second
night before settling to $0\farcs 9$ for the last hour.

The spectrograph was configured in the same manner for all of the
observations.  We used the 1200G grating with a groove spacing of
1200~mm$^{-1}$ and a blaze wavelength of 7760~\AA\@.  The grating was
tilted so that first-order light spanned about 6400--9000~\AA\ across
the CCD mosaic.  The central wavelength of each spectrum was
approximately 7800~\AA, but the exact spectral range depended on the
placement of the slit along the slitmask.  The slit widths were
$0\farcs 7$ for all slitmasks except for Leo~A, for which the slit
widths were $1\farcs 1$.  The resolving powers for the two slit widths
at 8500~\AA\ were $R \sim 7100$ and 4700, respectively.  We reduced
the data into sky-subtracted, one-dimensional spectra with the {\sc
  spec2d} software pipeline \citep{coo12,new13}.


\section{Velocity measurements}
\label{sec:vel}

We measured radial velocities by cross-correlating the observed
spectra with templates also observed with DEIMOS\@.  J.~Simon and
M.~Geha kindly provided the same templates they used to measure
velocities of red giants in ultra-faint dwarf galaxies \citep{sim07}.
The templates included red giants at a variety of metallicities as
well as a few K and M dwarfs.

Before cross-correlation, we normalized the observed spectrum.  We
fitted a B-spline with a breakpoint every 100~pixels to `continuum
regions' that \citet*{kir08} determined to be largely free of
absorption lines.  The continuum regions also excluded telluric
absorption.  The B-spline fit was weighted by the inverse variance of
each pixel.  For more details on the continuum division, please refer
to \citet[][sec.~3.4]{kir09}.

Following the same procedure as \citet{sim07}, we computed the
velocity of the maximum cross-correlation for each of the 16 template
spectra.  In contrast to \citeauthor{sim07}, we used only the red half
of the spectrum.  The DEIMOS focal plane is an eight-CCD mosaic.  The
CCDs are distributed in two rows of four detectors.  Each stellar
spectrum spans two CCDs.  The {\sc spec2d} reduction code computes
separate velocity solutions for the red and blue chips.  In order to
eliminate any systematic difference between the two wavelength
solutions, we used the half of the spectrum only from the red CCD to
compute the velocity.  This half of the spectrum includes the
Ca$\,${\sc ii} near-infrared triplet at 8498, 8542 and 8662~\AA\@.
These absorption lines dominate the cross-correlation.

We adopted the velocity corresponding to the template with the lowest
reduced $\chi^2$.  We shifted this velocity to the heliocentric frame
based on the position in the sky and the time of observation.  We
checked every spectrum by eye to ensure that the observed spectrum
lined up with the best-matching template spectrum.  In the rare cases
where the velocity was obviously wrong, we excluded certain regions of
the spectrum, such as the margins of the CCD where vignetting can
cause a sharp dip in flux, and we recomputed the velocity.

We estimated the Monte Carlo error on the velocity by resampling the
spectrum 1000 times.  In each Monte Carlo realisation, we added a
random value to the continuum-normalized flux in each pixel.  The
random value was chosen from a normal distribution with a standard
deviation equal to the estimated continuum-normalized variance on the
pixel flux.  We recomputed the velocity that maximized the
cross-correlation using the best-fitting template spectrum determined
above.

\citet{sim07} found that the Monte Carlo error is an incomplete
description of the total error on velocity.  From repeat measurements
of red giants in ultra-faint dwarf galaxies, they determined a
systematic error floor of 2.2~km~s$^{-1}$.  \citet{kir12,kir13a}
repeated this exercise with the isolated dIrr VV~124 and the
ultra-faint dSph Segue~2.  They determined error floors of
2.21~km~s$^{-1}$ and 1.95~km~s$^{-1}$.  Our sample contains very few
repeat measurements other than the VV~124 sample already analysed by
\citet{kir12}.  Therefore, we adopt a systematic error of
2.1~km~s$^{-1}$, which we added in quadrature to the random
uncertainties determined from the Monte Carlo resampling.  The
velocity dispersions for the isolated dwarf galaxies in this paper are
at least several times this systematic error.  Hence, the difference
between 2.21~km~s$^{-1}$ and 1.95~km~s$^{-1}$ is inconsequential.

\begin{figure}
 \centering
 \includegraphics[width=85mm]{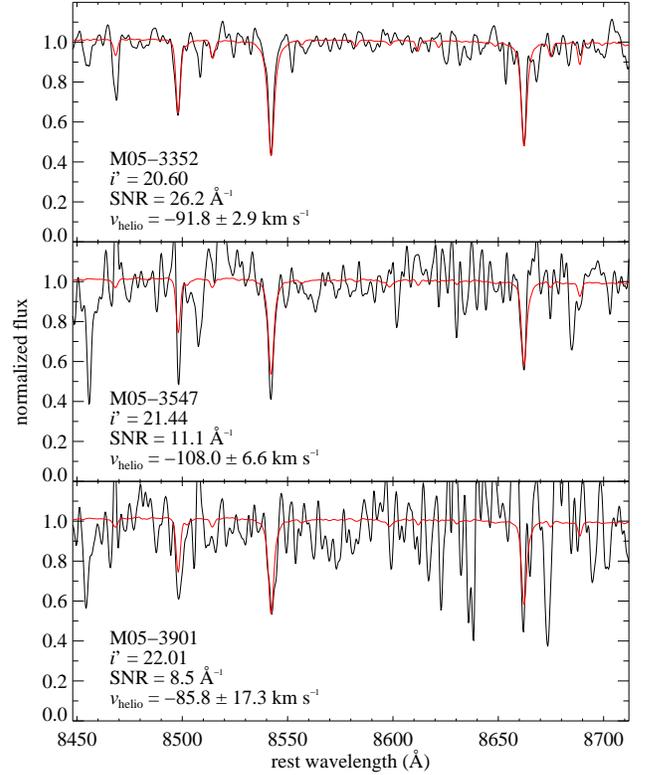}
 \caption{Examples of DEIMOS spectra of red giants in the Cetus dSph.
   The black lines show the DEIMOS spectra smoothed with a Gaussian
   with $\rmn{FWHM} = 1.6$~\AA\@.  The red lines show the best-fitting
   radial velocity template spectra.  The top spectrum has among the
   highest S/N in the Cetus sample, and the bottom has among the
   lowest S/N\@.\label{fig:spectra}}
\end{figure}

\begin{figure}
 \centering
 \includegraphics[width=85mm]{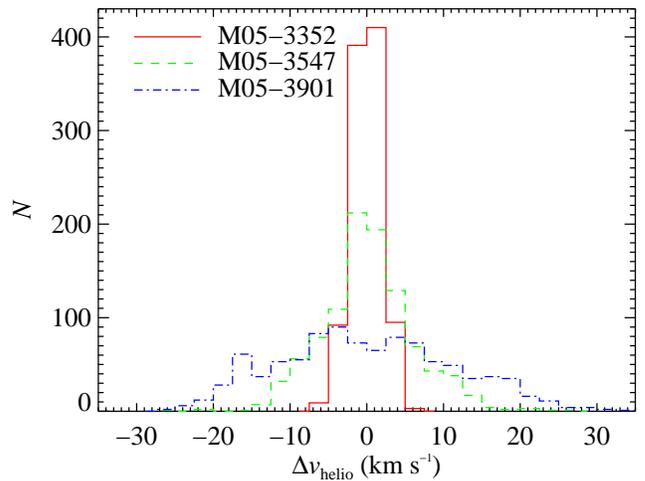}
 \caption{Probability distributions for the radial velocities of the
   spectra shown in Fig.~\ref{fig:spectra}.  The histograms show the
   distribution of the differences between the measured
   $v_{\rmn{helio}}$ and the velocities for 1000 Monte Carlo
   trials.\label{fig:mcmc}}
\end{figure}

\begin{table*}
\centering
\begin{minipage}{144mm}
\caption{Catalogue of velocities for individual stars.}
\label{tab:vel}
\begin{tabular}{@{}llcccr@{ }c@{ }lcc@{}}
\hline
Galaxy & Object Name & RA (J2000) & Dec (J2000) & S/N (\AA$^{-1}$) & \multicolumn{3}{c}{$v_{\rmn{helio}}$ (km~s$^{-1}$)} & Member? & Reason \\
\hline
IC 1613  & B07-44743  & 01 04 22.55 & $+$02 09 33.3 &  28.2 & $-254.2$ & $\pm$ &  2.6 & Y &      \\
IC 1613  & B07-44686  & 01 04 23.06 & $+$02 10 09.1 &  20.4 & $-237.1$ & $\pm$ &  3.0 & Y &      \\
IC 1613  & B07-41610  & 01 04 23.30 & $+$02 08 50.8 &  26.6 & $-263.9$ & $\pm$ &  2.6 & N & v    \\
IC 1613  & B07-44467  & 01 04 25.31 & $+$02 10 29.2 &  12.3 & $-240.4$ & $\pm$ &  5.6 & Y &      \\
IC 1613  & B07-44459  & 01 04 25.40 & $+$02 11 02.6 &  19.6 & $-231.5$ & $\pm$ &  3.2 & Y &      \\
IC 1613  & B07-37893  & 01 04 25.53 & $+$02 08 39.0 &  19.9 & $-242.3$ & $\pm$ &  2.7 & Y &      \\
IC 1613  & B07-47494  & 01 04 26.16 & $+$02 11 47.9 &  13.3 & $-244.0$ & $\pm$ &  4.3 & Y &      \\
IC 1613  & B07-41149  & 01 04 26.26 & $+$02 10 22.3 &  34.3 & $-252.5$ & $\pm$ &  2.3 & Y &      \\
IC 1613  & B07-47458  & 01 04 26.98 & $+$02 11 50.5 &  13.3 & $-243.8$ & $\pm$ &  3.9 & Y &      \\
IC 1613  & B07-41044  & 01 04 27.32 & $+$02 10 35.1 &  20.3 & $-249.0$ & $\pm$ &  3.0 & Y &      \\
\hline
\end{tabular}

\medskip
`Reason' indicates reasons for non-membership.  v: Radial velocity non-member.  Na: Na$\,${\sc i}~$\lambda$8190 EW exceeds 1~\AA\@.  This table has 862 rows, of which only the first 10 rows are reproduced here.
\end{minipage}
\end{table*}

Fig.~\ref{fig:spectra} shows some example spectra at a range of S/N\@.
The S/N was calculated as the median absolute deviation from the
continuum in the continuum regions.  (See sec.~2.3 of
\citeauthor{kir12}\ \citeyear{kir12} for more details.)  The figure
also shows the best-matching template spectra.  The Ca triplet line
strengths do not necessarily match between the two spectra, but the
spectra align in wavelength.

The S/N is quite low in the lowest quality spectra, such as M05--3901
in Fig.~\ref{fig:spectra}.  However, cross-correlation can identify
patterns in spectral features that are not easy to see by eye.  It is
nonetheless worthwhile to validate velocity measurements for cases
that seem questionable.  Fig.~\ref{fig:mcmc} shows the results of the
1000 Monte Carlo trials of measuring $v_{\rmn{helio}}$ for the three
spectra shown in Fig.~\ref{fig:spectra}.  Even in the case of the
lowest S/N, the velocities are distributed roughly normally about the
measured velocity.  If the spectrum contained no useful information,
the velocities from the 1000 trials would be distributed uniformly
over the velocity range that the cross-correlation searched.
Furthermore, there is only one peak in each probability distribution.
Therefore, the cross-correlation found a unique velocity for each
star.

It was not possible to measure velocities of some stars because the
S/N was too low.  If we could identify no clear Ca triplet in the
spectrum, then we marked the spectrum as unusable.  We also excluded
from our sample all stars with velocity errors in excess of
30~km~s$^{-1}$.  Stars with unusable spectra are marked as crosses in
Figs.~\ref{fig:cetcoords} and \ref{fig:cetcmd}.

Table~\ref{tab:vel} gives the celestial coordinates, S/N ratios,
heliocentric radial velocities and velocity uncertainties (random and
systematic errors added in quadrature) for each star in all seven
dwarf galaxies.  It also includes information about our membership
determinations for each star (see Sec.~\ref{sec:membership}).

\begin{figure}
 \centering
 \includegraphics[width=85mm]{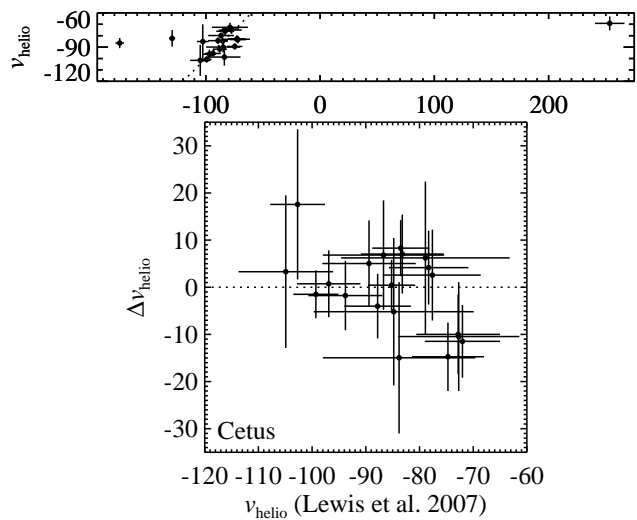}
 \caption{Comparison of radial velocities of the stars in common
   between our work and that of \protect\citet{lew07}.  The top panel
   shows all \nlewmatch~stars in common.  The bottom panel shows the
   \nlewmatchon~stars in the range $-120 < v_{\rmn{helio}} <
   -60$~km~s$^{-1}$.  The dotted lines indicate
   equality.\label{fig:lew07}}
\end{figure}

\subsection{Comparison to previous measurements}
\label{sec:lew07}

\citet{lew07} measured radial velocities for \nlew\ stars in the
vicinity of Cetus.  Fig.~\ref{fig:lew07} presents the comparison of
the radial velocity measurements of the \nlewmatch~stars that overlap
with our sample.  The velocity differences, $\Delta v_{\rmn{helio}}$,
are distributed as expected based on the estimated uncertainties for
\nlewmatchon~stars.  The remaining three stars are highly discrepant
between the two samples.  We would not have considered these stars
members of Cetus based on \citeauthor{lew07}'s velocities, but our
velocities are within our membership cuts.  We confirmed that the
radial velocity template spectra line up with our observed spectra
with our velocities but not with \citeauthor{lew07}'s velocities.  The
differences in velocities between the two studies is probably due to
the longer exposure times and higher S/N of our spectra.


\section{Membership}
\label{sec:membership}

Not every star that we targeted is a member of its respective galaxy.
We assigned binary (yes or no) membership to each star based on its
position in the colour--magnitude diagram, the equivalent width (EW) of
its Na$\,${\sc i}~$\lambda$8190 doublet and its radial velocity.  We
used only member stars to compute average radial velocities and
velocity dispersions of the dwarf galaxies.

\subsection{Colour--magnitude diagram}

The selection in the colour--magnitude diagram was accomplished in the
slitmask design before the spectra were obtained.  Only stars with
reasonable colours and magnitudes for red giants were allowed.
Fig.~\ref{fig:cetcmd} of this paper, fig.~1 of \citet{kir12} and
figs.~2--6 of \citet{kir13b} show that all of the stars deemed to be
members are indeed on the red giant branch.

\subsection{Na doublet}

The Na$\,${\sc i} doublet at 8183 and 8195~\AA\ is a good indicator of
surface gravity \citep{spi71,coh78,sch97,gil06}.  Dwarf stars, which
have high surface gravities, have strong Na doublets.  \citet{kir12}
showed that the combined EW of the doublet exceeds 1~\AA\ only in
stars with surface gravities $\log g > 4.5$ as long as $\rmn{[Na/H]}
\le 0$, which is a reasonable assumption for these metal-poor galaxies
\citep{kir13b}.

We computed EWs for each line in the doublet by fitting Gaussian or
Lorentzian profiles.  For weak doublets, Gaussians were better fits.
For very strong doublets in dwarf stars, Lorentzian profiles better
matched the damping wings of the absorption lines.  In most stars, the
doublet was not detectable above the noise.  For those stars where it
was possible to measure EWs, we also computed errors by Monte Carlo
resampling of the spectra in the same manner used to compute velocity
errors (see Sec.~\ref{sec:vel}).

We ruled as non-members those stars where the combined EW of the two
lines exceeded 1~\AA, even after accounting for the error in the EW
measurements.  In other words, for a star to be counted as a
non-member, its Na$\,${\sc i} EW needed to exceed 1~\AA\ by at least
$1\sigma$.  Stars ruled as non-members on the basis of Na$\,${\sc i}
EW are indicated by `Na' in the last column of Table~\ref{tab:vel}.

\subsection{Radial velocity}

Some stars have radial velocities inconsistent with the radial
velocity of the galaxy.  We interpret these interlopers as foreground
stars in the Milky Way.  We set the exclusion limits at 2.58 times the
velocity dispersion (determined in Sec.~\ref{sec:disp}).  This range
includes 99~per cent of the member stars assuming a normal velocity
distribution.  It excludes non-members at the cost of also excluding
1~per cent of members.

However, we retained stars with velocity errors within $1\sigma$ of
the allowed range of velocities.  For example, we measured the mean
velocity of IC~1613 to be $\langle v_{\rmn{helio}} \rangle =
\icosotvmean$~km~s$^{-1}$ with a velocity dispersion of $\sigma_v =
\icosotsigmav$~km~s$^{-1}$.  Therefore, the range of allowed
velocities is $\icosotvlow < v_{\rmn{helio}} <
\icosotvhigh$~km~s$^{-1}$.  Star B07-56088 has a radial velocity of
$\icosotstarv \pm \icosotstarverr$~km~s$^{-1}$.  Although the velocity
of this star is outside of the membership limits, we still counted it
as a member because its $1\sigma$ error bar reaches the upper range of
velocities that qualify for membership in IC~1613.

The membership list affects the measurements of $\langle
v_{\rmn{helio}} \rangle$ and $\sigma_v$, but the membership criteria
depend on those measurements.  Therefore, the membership determination
was iterative.  We started with guesses at $\langle v_{\rmn{helio}}
\rangle$ and $\sigma_v$, determined from fitting a Gaussian to a
velocity histogram.  Then, we determined membership for each star
based on radial velocity, and we calculated $\langle v_{\rmn{helio}}
\rangle$ and $\sigma_v$ following the procedure described in
Sec.~\ref{sec:disp}.  We performed a new membership cut based on
these new values, and we repeated the process until the membership
list did not change from one iteration to the next.

Stars that were excluded on the basis of radial velocity are indicated
by `v' in the last column of Table~\ref{tab:vel}.  Stars ruled as
non-members from either Na$\,${\sc i} EW or radial velocity are shown
as hollow circles in Figs.~\ref{fig:cetcoords} and \ref{fig:cetcmd}.


\section{Velocity dispersions}
\label{sec:disp}

\begin{figure*}
 \centering
 \includegraphics[width=160mm]{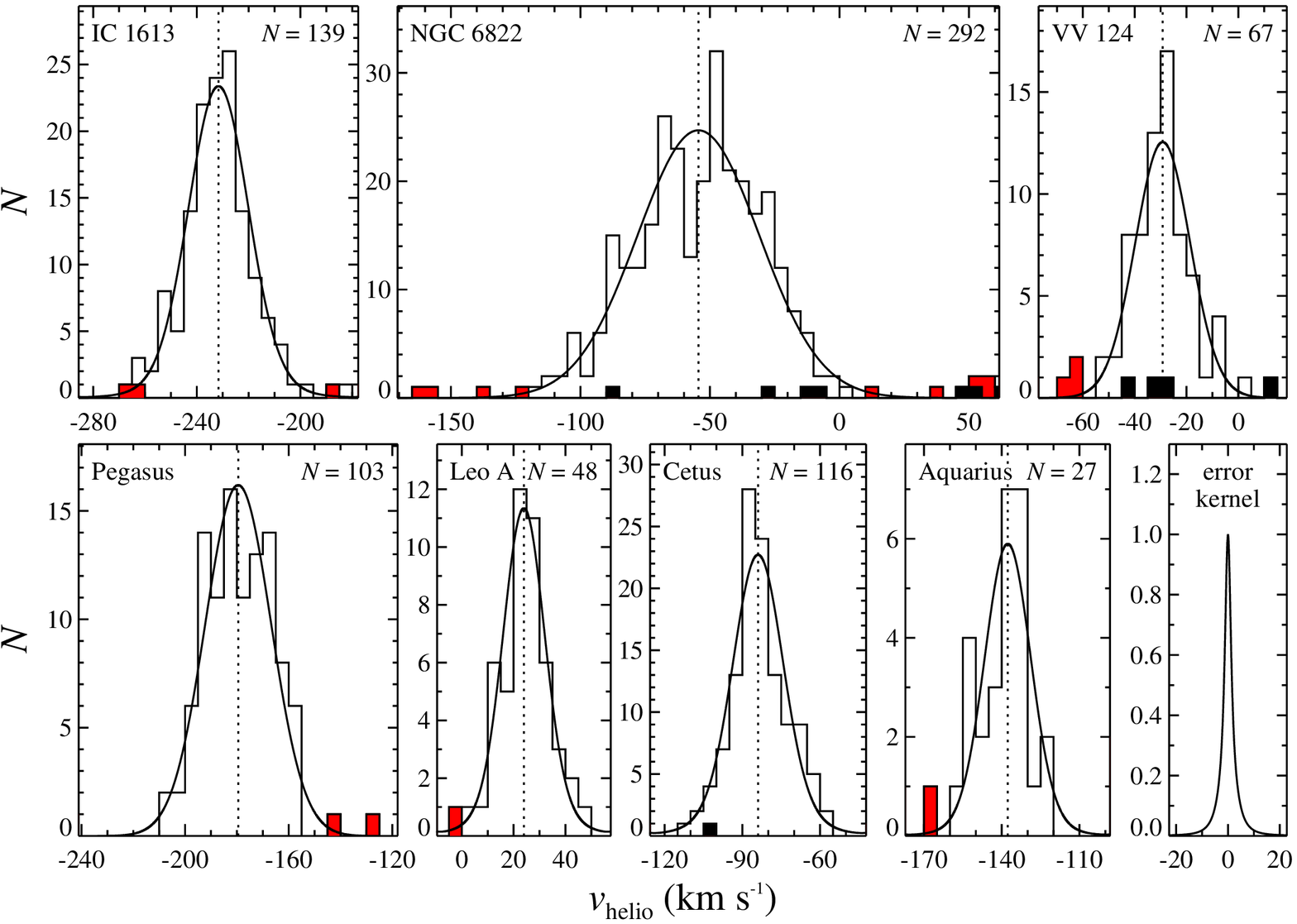}
 \caption{Heliocentric radial velocity distributions.  Unshaded
   regions of the histograms contain spectroscopically confirmed
   members.  Black and red shaded regions indicate spectroscopic
   non-members from Na$\,${\sc i}~$\lambda$8190 EW and radial
   velocity, respectively.  The vertical dotted lines indicate
   $\langle v_{\rmn{helio}} \rangle$.  The solid curve is a Gaussian
   with the measured $\sigma_v$, convolved with the error kernel,
   which is shown in the bottom right panel.  The bins are
   5~km~s$^{-1}$ wide.  The number of member stars is indicated in the
   upper right of each panel.  The rotation curve has been removed
   from Pegasus (see Sec.~\ref{sec:pegrot}).\label{fig:vhist}}
\end{figure*}

We measured $\langle v_{\rmn{helio}} \rangle$ and $\sigma_v$ with a
Monte Carlo Markov chain (MCMC)\@.  Following the procedure for
measuring velocity dispersions introduced by \citet{wal06a}, we
maximized the likelihood ($L$) that the values of $\langle
v_{\rmn{helio}} \rangle$ and $\sigma_v$ accurately described the
observed velocity distribution of member stars, accounting for
uncertainties in the individual velocity measurements.  The logarithm
of the likelihood is

\begin{eqnarray}
\nonumber \log L &=& \frac{N \log(2 \pi)}{2} + \frac{1}{2} \sum_i^N \left(\log((\delta v_r)_i^2 + \sigma_v^2\right) \\
& & + \frac{1}{2} \sum_i^N \left(\frac{((v_{\rmn{helio}})_i - \langle v_{\rmn{helio}} \rangle)^2}{(\delta v_r)_i^2 + \sigma_v^2}\right) \label{eq:l}
\end{eqnarray}

\noindent
where $N$ is the number of member stars and $(v_{\rmn{helio}})_i$ and
$(\delta v_r)_i$ are the velocity and error for star $i$.

The MCMC had a length of $10^7$ trials.  We implemented the
Metropolis--Hastings algorithm with normally distributed proposal
densities.  The standard deviation of both the $\langle
v_{\rmn{helio}} \rangle$ and $\sigma_v$ proposal densities was
5~km~s$^{-1}$.  The values for each iteration ($j$) were perturbed
from the previous iteration ($j-1$) according to these proposal
densities.  If the likelihood increased or if $\exp (L_j - L_{j-1})$
was greater than a random number selected from a uniform distribution
between 0 and 1, then the new values of $\langle v_{\rmn{helio}}
\rangle$ and $\sigma_v$ from iteration $j$ were accepted.  Otherwise,
they were discarded and the next iteration began with the original
values of $\langle v_{\rmn{helio}} \rangle$ and $\sigma_v$ from
iteration $j-1$.

\begin{table*}
\centering
\begin{minipage}{176mm}
\caption{Structural and dynamical quantities for Local Group galaxies with $10^5 < L_V/\lsun < 2 \times 10^8$.}
\label{tab:summary}
\begin{tabular}{@{}lr@{ }lr@{ }lr@{ }lrr@{ }lr@{ }lr@{ }lr@{ }l@{}}
\hline
Galaxy & \multicolumn{2}{c}{Distance} & \multicolumn{2}{c}{$L_V$} & \multicolumn{2}{c}{$r_h$} & $N$ & \multicolumn{2}{c}{$\langle v_{\rmn{helio}} \rangle$} & \multicolumn{2}{c}{$\sigma_v$} & \multicolumn{2}{c}{$M_{1/2}$} & \multicolumn{2}{c}{$(M/L_V)_{1/2}$} \\
 & \multicolumn{2}{c}{(kpc)} & \multicolumn{2}{c}{($10^6~\lsun$)} & \multicolumn{2}{c}{(pc)} & & \multicolumn{2}{c}{(km~s$^{-1}$)} & \multicolumn{2}{c}{(km~s$^{-1}$)} & \multicolumn{2}{c}{($10^6~\msun$)} & \multicolumn{2}{c}{($\msun~\lsun^{-1}$)} \\
\hline
\multicolumn{16}{c}{Isolated (this work)} \\
\hline
IC 1613  &           758 & $\pm$   4$^{ c}$ &                  100 & $^{+ 20}_{- 10}$$^{ d}$ &          1040 & $\pm$  65$^{ e}$ & 139 &     $-231.6$ & $\pm$  1.2 &   10.8 & $^{+1.0}_{-0.9}$$^{  }$ &                         110 & $\pm$  20 &                   2.2 & $\pm$ 0.5 \\
NGC 6822 &           459 & $\pm$   8$^{ f}$ &                           98 & $\pm$ 18$^{ g}$ &           478 & $\pm$  28$^{ e}$ & 292 &     $ -54.5$ & $\pm$  1.7 &          23.2 & $\pm$ 1.2$^{  }$ &                         240 & $\pm$  30 &                   4.9 & $\pm$ 1.1 \\
VV 124   &          1361 & $\pm$  25$^{ h}$ &                  9.0 & $^{+1.8}_{-1.5}$$^{ i}$ &           272 & $\pm$  27$^{ i}$ &  67 &     $ -29.2$ & $\pm$  1.6 &    9.6 & $^{+1.3}_{-1.2}$$^{  }$ &                     23 & $^{+ 7}_{- 6}$ &            5.2 & $^{+1.8}_{-1.7}$ \\
Pegasus  &           920 & $\pm$  29$^{ j}$ &                  6.6 & $^{+1.4}_{-1.2}$$^{ d}$ &           695 & $\pm$  37$^{ e}$ & 103 &     $-179.5$ & $\pm$  1.5 &   12.3 & $^{+1.2}_{-1.1}$$^{  }$ &                         130 & $\pm$  10$^a$ &                     39 & $\pm$  8$^a$ \\
Leo A    &           787 & $\pm$   4$^{ k}$ &                  6.0 & $^{+1.4}_{-1.2}$$^{ d}$ &           354 & $\pm$  19$^{ e}$ &  48 &     $  24.0$ & $\pm$  1.5 &    6.7 & $^{+1.4}_{-1.2}$$^{  }$ &                     15 & $^{+ 6}_{- 5}$ &            5.0 & $^{+2.3}_{-2.1}$ \\
Cetus    &           779 & $\pm$  43$^{ l}$ &                         3.0 & $\pm$ 0.6$^{ m}$ &           612 & $\pm$  38$^{ m}$ & 116 &     $ -83.9$ & $\pm$  1.2 &           8.3 & $\pm$ 1.0$^{  }$ &                     40 & $^{+10}_{- 9}$ &                     27 & $\pm$  9 \\
Aquarius &          1071 & $\pm$  39$^{ j}$ &                         1.2 & $\pm$ 0.1$^{ n}$ &           342 & $\pm$  15$^{ n}$ &  27 &     $-137.7$ & $\pm$  2.1 &    7.9 & $^{+1.9}_{-1.6}$$^{  }$ &                     20 & $^{+10}_{- 8}$ &               32 & $^{+16}_{-14}$ \\
\hline
\multicolumn{16}{c}{Isolated (literature)} \\
\hline
WLM                       &           933 & $\pm$  34$^{ j}$ &                           43 & $\pm$  5$^{ d}$ &          1569 & $\pm$  74$^{ o}$ & &     $-130.0$ & $\pm$  1.0 &          17.0 & $\pm$ 1.0$^{ o}$ &                         630 & $\pm$  30$^{a,b}$ &                     30 & $\pm$  4$^{a,b}$ \\
Tucana                    &           887 & $\pm$  49$^{ l}$ &                       0.59 & $\pm$ 0.12$^{ p}$ &           209 & $\pm$  34$^{ p}$ & &     $ 194.0$ & $\pm$  4.3 &   15.8 & $^{+4.1}_{-3.1}$$^{ q}$ &                           71 & $\pm$ 12$^{a}$ &                   240 & $\pm$  60$^{a}$ \\
Leo T                     &           398 & $\pm$  36$^{ r}$ &                       0.14 & $\pm$ 0.04$^{ r}$ &           114 & $\pm$  12$^{ r}$ & &     $  38.1$ & $\pm$  2.0 &           7.5 & $\pm$ 1.6$^{ s}$ &                         6.0 & $\pm$ 2.6 &                     89 & $\pm$ 46 \\
\hline
\multicolumn{16}{c}{Milky Way} \\
\hline
Sagittarius               &            26 & $\pm$   1$^{ t}$ &                           21 & $\pm$  6$^{ u}$ &          1551 & $\pm$ 118$^{ u}$ & &     $ 140.0$ & $\pm$  2.0 &           9.9 & $\pm$ 0.7$^{ v}$ &                         140 & $\pm$  20 &                     14 & $\pm$  5 \\
Fornax                    &           147 & $\pm$   9$^{ w}$ &                           20 & $\pm$  6$^{ x}$ &           710 & $\pm$  70$^{ x}$ & &     $  55.1$ & $\pm$  0.3 &          11.8 & $\pm$ 0.2$^{ y}$ &                           91 & $\pm$ 10 &                   9.0 & $\pm$ 2.7 \\
Leo I                     &           253 & $\pm$  15$^{ z}$ &                         5.6 & $\pm$ 1.5$^{ x}$ &           250 & $\pm$  26$^{ x}$ & &     $ 282.8$ & $\pm$  0.6 &           8.8 & $\pm$ 0.5$^{aa}$ &                           18 & $\pm$  3 &            6.5 & $^{+2.1}_{-2.0}$ \\
Sculptor                  &            85 & $\pm$   4$^{bb}$ &                         2.3 & $\pm$ 1.1$^{ x}$ &           282 & $\pm$  42$^{ x}$ & &     $ 111.4$ & $\pm$  0.3 &           9.2 & $\pm$ 0.2$^{ y}$ &                           22 & $\pm$  4 &                     20 & $\pm$ 10 \\
Leo II                    &           233 & $\pm$  13$^{cc}$ &                       0.75 & $\pm$ 0.21$^{ x}$ &           176 & $\pm$  42$^{ x}$ & &     $  78.0$ & $\pm$  0.7 &           6.6 & $\pm$ 0.7$^{dd}$ &                         7.1 & $\pm$ 2.3 &                     19 & $\pm$  8 \\
Sextans                   &            85 & $\pm$   3$^{ee}$ &                       0.44 & $\pm$ 0.20$^{ x}$ &           694 & $\pm$  43$^{ x}$ & &     $ 224.2$ & $\pm$  0.5 &           7.9 & $\pm$ 0.4$^{ y}$ &                     40 & $^{+ 4}_{- 5}$ &                   180 & $\pm$  90 \\
Carina                    &           106 & $\pm$   7$^{ w}$ &                       0.37 & $\pm$ 0.17$^{ x}$ &           254 & $\pm$  40$^{ x}$ & &     $ 223.1$ & $\pm$  0.4 &           6.7 & $\pm$ 0.3$^{ y}$ &                           11 & $\pm$  2 &                     57 & $\pm$ 28 \\
Ursa Minor                &            75 & $\pm$   3$^{ff}$ &                       0.27 & $\pm$ 0.13$^{ x}$ &           180 & $\pm$  27$^{ x}$ & &     $-246.9$ & $\pm$  0.8 &           9.5 & $\pm$ 1.2$^{gg}$ &                           15 & $\pm$  4 &                   110 & $\pm$  60 \\
Draco                     &            75 & $\pm$   5$^{hh}$ &                       0.27 & $\pm$ 0.05$^{ii}$ &    220 & $^{+ 16}_{- 15}$$^{ii}$ & &     $-291.0$ & $\pm$  0.7 &           9.1 & $\pm$ 1.2$^{dd}$ &                           17 & $\pm$  5 &                   130 & $\pm$  40 \\
Can.\ Ven.\ I             &           217 & $\pm$  23$^{jj}$ &               0.24 & $^{+0.05}_{-0.07}$$^{ii}$ &           564 & $\pm$  64$^{ii}$ & &     $  30.9$ & $\pm$  0.6 &           7.6 & $\pm$ 0.4$^{ s}$ &                           30 & $\pm$  5 &                   260 & $\pm$  80 \\
\hline
\multicolumn{16}{c}{M31} \\
\hline
NGC 185                   &    619 & $^{+ 19}_{- 17}$$^{kk}$ &                     68 & $^{+ 9}_{- 8}$$^{ d}$ &    459 & $^{+ 91}_{- 90}$$^{ d}$ & &     $-203.8$ & $\pm$  1.1 &          24.0 & $\pm$ 1.0$^{ll}$ &                         290 & $\pm$  60$^{a}$ &                   8.7 & $\pm$ 2.0$^{a}$ \\
NGC 147                   &           711 & $\pm$  19$^{kk}$ &                     62 & $^{+ 8}_{- 7}$$^{ d}$ &           655 & $\pm$ 104$^{ d}$ & &     $-193.1$ & $\pm$  0.8 &          16.0 & $\pm$ 1.0$^{ll}$ &                         240 & $\pm$  40$^{a}$ &                   7.9 & $\pm$ 1.6$^{a}$ \\
And VII                   &           762 & $\pm$  35$^{ j}$ &                           17 & $\pm$  5$^{ m}$ &           731 & $\pm$  36$^{ m}$ & &     $-307.2$ & $\pm$  1.3 &          13.0 & $\pm$ 1.0$^{mm}$ &                         110 & $\pm$  20 &                     13 & $\pm$  4 \\
And II                    &           630 & $\pm$  14$^{kk}$ &                         8.6 & $\pm$ 1.6$^{ m}$ &          1027 & $\pm$  30$^{ m}$ & &     $-192.4$ & $\pm$  0.5 &           7.8 & $\pm$ 1.1$^{nn}$ &                         110 & $\pm$   0$^{a}$ &                     27 & $\pm$  5$^{a}$ \\
And I                     &           727 & $\pm$  16$^{kk}$ &                         4.5 & $\pm$ 0.5$^{ m}$ &           592 & $\pm$  25$^{ m}$ & &     $-376.3$ & $\pm$  2.2 &          10.2 & $\pm$ 1.9$^{mm}$ &                           57 & $\pm$ 22 &                     26 & $\pm$ 10 \\
And VI                    &           783 & $\pm$  25$^{ j}$ &                         3.4 & $\pm$ 0.7$^{ m}$ &           410 & $\pm$  20$^{ m}$ & &     $-339.8$ & $\pm$  1.8 &   12.4 & $^{+1.5}_{-1.3}$$^{oo}$ &                     59 & $^{+14}_{-13}$ &               35 & $^{+11}_{-10}$ \\
LGS 3                     &           769 & $\pm$  24$^{ j}$ &                         1.1 & $\pm$ 0.1$^{pp}$ &           469 & $\pm$  47$^{pp}$ & &     $-282.2$ & $\pm$  3.5 &    7.9 & $^{+5.3}_{-2.9}$$^{qq}$ &                     27 & $^{+37}_{-20}$ &               49 & $^{+66}_{-37}$ \\
And XXIII                 &    748 & $^{+ 31}_{- 20}$$^{kk}$ &                       0.97 & $\pm$ 0.45$^{rr}$ &   1001 & $^{+ 60}_{- 51}$$^{rr}$ & &     $-237.7$ & $\pm$  1.2 &           7.1 & $\pm$ 1.0$^{oo}$ &                     47 & $^{+14}_{-13}$ &                     96 & $\pm$ 52 \\
And III                   &    724 & $^{+ 16}_{- 23}$$^{kk}$ &                       0.93 & $\pm$ 0.26$^{ m}$ &    337 & $^{+ 19}_{- 20}$$^{ m}$ & &     $-344.3$ & $\pm$  1.7 &           9.3 & $\pm$ 1.4$^{mm}$ &                           27 & $\pm$  8 &                     58 & $\pm$ 24 \\
And XXI                   &    827 & $^{+ 22}_{- 26}$$^{kk}$ &                       0.78 & $\pm$ 0.43$^{ss}$ &    842 & $^{+ 75}_{- 77}$$^{ss}$ & &     $-361.4$ & $\pm$  5.8 &           7.2 & $\pm$ 5.5$^{mm}$ &                                $<150$ & &                          $<390$ & \\
And XXV                   &    734 & $^{+ 23}_{- 71}$$^{kk}$ &                       0.53 & $\pm$ 0.24$^{rr}$ &    640 & $^{+ 47}_{- 75}$$^{rr}$ & &     $-107.8$ & $\pm$  1.0 &    3.0 & $^{+1.2}_{-1.1}$$^{oo}$ &                  5.4 & $^{+4.3}_{-4.0}$ &               20 & $^{+19}_{-18}$ \\
And V                     &    741 & $^{+ 20}_{- 23}$$^{kk}$ &                       0.52 & $\pm$ 0.10$^{ m}$ &    280 & $^{+ 19}_{- 20}$$^{ m}$ & &     $-397.3$ & $\pm$  1.5 &          10.5 & $\pm$ 1.1$^{mm}$ &                           29 & $\pm$  6 &                   110 & $\pm$  30 \\
And XV                    &    625 & $^{+ 74}_{- 34}$$^{kk}$ &                       0.49 & $\pm$ 0.18$^{tt}$ &    220 & $^{+ 27}_{- 15}$$^{tt}$ & &     $-323.0$ & $\pm$  1.4 &           4.0 & $\pm$ 1.4$^{mm}$ &                         3.3 & $\pm$ 2.3 &                     13 & $\pm$ 11 \\
And XIX                   &    820 & $^{+ 30}_{-162}$$^{kk}$ &                       0.45 & $\pm$ 0.25$^{uu}$ &   1479 & $^{+ 59}_{-293}$$^{uu}$ & &     $-111.6$ & $\pm$  1.5 &    4.7 & $^{+1.6}_{-1.4}$$^{oo}$ &                     30 & $^{+21}_{-19}$ &            140 & $^{+120}_{-110}$ \\
And XVI                   &    476 & $^{+ 41}_{- 30}$$^{kk}$ &                       0.41 & $\pm$ 0.15$^{tt}$ &    123 & $^{+ 12}_{- 10}$$^{tt}$ & &     $-367.3$ & $\pm$  2.8 &           3.8 & $\pm$ 2.9$^{mm}$ &                                $<5.9$ & &                           $<30$ & \\
And XXVII                 &   1253 & $^{+ 40}_{-594}$$^{kk}$ &                       0.28 & $\pm$ 0.13$^{rr}$ &    656 & $^{+111}_{-329}$$^{rr}$ & &     $-539.6$ & $\pm$  4.6 &   14.8 & $^{+4.3}_{-3.1}$$^{oo}$ &                  130 & $^{+ 80}_{- 90}$ &            940 & $^{+720}_{-760}$ \\
And XVII                  &    727 & $^{+ 36}_{- 26}$$^{kk}$ &                       0.24 & $\pm$ 0.07$^{vv}$ &    262 & $^{+ 21}_{- 19}$$^{vv}$ & &     $-251.6$ & $\pm$  1.9 &    2.9 & $^{+2.2}_{-1.9}$$^{oo}$ &                                $<7.3$ & &                           $<62$ & \\
And XIV                   &    794 & $^{+ 21}_{-204}$$^{kk}$ &               0.24 & $^{+0.23}_{-0.09}$$^{ww}$ &    392 & $^{+185}_{-210}$$^{ww}$ & &     $-480.6$ & $\pm$  1.2 &           5.3 & $\pm$ 1.0$^{mm}$ &                     10 & $^{+ 6}_{- 7}$ &               87 & $^{+71}_{-74}$ \\
And XXVIII                &    660 & $^{+152}_{- 60}$$^{xx}$ &               0.21 & $^{+0.20}_{-0.08}$$^{xx}$ &    213 & $^{+ 63}_{- 44}$$^{xx}$ & &     $-331.1$ & $\pm$  1.8 &           4.9 & $\pm$ 1.6$^{yy}$ &                  4.8 & $^{+3.4}_{-3.3}$ &               44 & $^{+43}_{-42}$ \\
And XXIX                  &           731 & $\pm$  74$^{zz}$ &                       0.18 & $\pm$ 0.07$^{zz}$ &           361 & $\pm$  56$^{zz}$ & &     $-194.4$ & $\pm$  1.5 &           5.7 & $\pm$ 1.2$^{yy}$ &                           11 & $\pm$  5 &                   120 & $\pm$  70 \\
And IX                    &    599 & $^{+ 85}_{- 22}$$^{kk}$ &               0.15 & $^{+0.01}_{-0.05}$$^{ab}$ &    432 & $^{+ 64}_{- 87}$$^{ab}$ & &     $-209.4$ & $\pm$  2.5 &          10.9 & $\pm$ 2.0$^{mm}$ &                     48 & $^{+19}_{-20}$ &            640 & $^{+290}_{-310}$ \\
And XXX                   &    682 & $^{+ 31}_{- 81}$$^{kk}$ &               0.13 & $^{+0.06}_{-0.03}$$^{oo}$ &    267 & $^{+ 26}_{- 48}$$^{oo}$ & &     $-139.8$ & $\pm$  6.3 &   11.8 & $^{+7.7}_{-4.7}$$^{oo}$ &                     35 & $^{+45}_{-28}$ &            530 & $^{+710}_{-460}$ \\
\hline
\end{tabular}

\medskip
The 2-D, projected half-light radius ($r_h$) is related to the 3-D, de-projected half-light radius ($r_{1/2}$) by $r_{1/2} \simeq \frac{4}{3}r_h$.  The last two columns give the mass and mass-to-light ratio within $r_{1/2}$.  $L_V$, $M_{1/2}$ and $(M/L_V)_{1/2}$ are quoted with two significant digits.  Where the error bars include zero, $2\sigma$ upper limits are given. \\
$^a$$M_{1/2}$ and $(M/L_V)_{1/2}$ include a contribution from rotation (see Sec.~\ref{sec:pegrot}). \\
$^b$\citet{lea12} calculated $M_{1/2} = (4.3 \pm 0.3) \times 10^8~\msun$ based on a sophisticated dynamical model that is more appropriate for a rotating system than our modification to \citeauthor{wol10}'s (\citeyear{wol10}) formula for $M_{1/2}$ (Sec.~\ref{sec:pegrot}).  \citeauthor{lea12}'s value of $M_{1/2}$ implies $(M/L_V)_{1/2} = 20.1 \pm 2.8~\msun~\lsun^{-1}$. \\
\begin{tabular}{@{}llll@{}}
$^{c}$\protect\citet{ber10} & $^{d}$\protect\citet{dev91} & $^{e}$\protect\citet{hun06} & $^{f}$\protect\citet{gie06} \\
\end{tabular}
\end{minipage}
\end{table*}
 
\begin{table*}
\centering
\begin{minipage}{176mm}
\contcaption{}
\begin{tabular}{@{}lll@{}}
$^{g}$\protect\citet{dal07} & $^{h}$\protect\citet{jac11} & $^{i}$\protect\citet{bel11} \\
$^{j}$\protect\citet{mcc05} & $^{k}$\protect\citet{ber13} & $^{l}$\protect\citet{ber09} \\
$^{m}$\protect\citet{mcc06} & $^{n}$\protect\citet{mccetal06} & $^{o}$\protect\citet{lea12} \\
$^{p}$\protect\citet{sav96} & $^{q}$\protect\citet{fra09} & $^{r}$\protect\citet{dej08} \\
$^{s}$\protect\citet{sim07} & $^{t}$\protect\citet{mon04} & $^{u}$\protect\citet{maj03} \\
$^{v}$\protect\citet{fri12} & $^{w}$\protect\citet{pie09} & $^{x}$\protect\citet{irw95} \\
$^{y}$\protect\citet{wal09} & $^{z}$\protect\citet{bel04} & $^{aa}$\protect\citet{mat08} \\
$^{bb}$\protect\citet{pie08} & $^{cc}$\protect\citet{bel05} & $^{dd}$\protect\citet{wal07} \\
$^{ee}$\protect\citet{lee09} & $^{ff}$\protect\citet{car02} & $^{gg}$\protect\citet{waletal09} \\
$^{hh}$\protect\citet{bon04} & $^{ii}$\protect\citet{mar08} & $^{jj}$\protect\citet{maretal08} \\
$^{kk}$\protect\citet{con12} & $^{ll}$\protect\citet{geh10} & $^{mm}$\protect\citet{tol12} \\
$^{nn}$\protect\citet{ho12} & $^{oo}$\protect\citet{col13} & $^{pp}$\protect\citet{lee95} \\
$^{qq}$\protect\citet{coo99} & $^{rr}$\protect\citet{ric11} & $^{ss}$\protect\citet{mar09} \\
$^{tt}$\protect\citet{iba07} & $^{uu}$\protect\citet{mcc08} & $^{vv}$\protect\citet{bra11} \\
$^{ww}$\protect\citet{maj07} & $^{xx}$\protect\citet{sla11} & $^{yy}$\protect\citet{tol13} \\
$^{zz}$\protect\citet{bell11} & $^{ab}$\protect\citet{col10} \\
\end{tabular}
\end{minipage}
\end{table*}

The final values of $\langle v_{\rmn{helio}} \rangle$ and $\sigma_v$
were set to be the mean values of the successful links in the MCMC\@.
The asymmetric $1\sigma$ confidence intervals were determined from the
values that enclosed 68.3~per cent of the successful MCMC links.
Table~\ref{tab:summary} gives these values for each of the seven dwarf
galaxies in our sample, along with the distances, $V$-band
luminosities, and half-light radii.  The table lists the original
sources for every measurement, but we adopted \citeauthor{mcc12}'s
(\citeyear{mcc12}) conversions from various scale radii to half-light
radii in cases where the original sources did not quote half-light
radii.  The table also gives these values from the literature for all
of the other Local Group galaxies with $10^5 < L_V/\lsun < 2 \times
10^8$ whose velocity dispersions have been measured.  The luminosity
cut restricts the satellite galaxies to about the same stellar mass
range as the isolated galaxies.  For the galaxies in our sample, the
table indicates the number of member stars ($N$).  The table also
gives the total dynamical mass enclosed within the de-projected,
three-dimensional half-light radius ($M_{1/2}$) from the formula of
\citet{wol10}.  The 3-D half-light radius, $r_{1/2}$, is well
approximated by $\frac{4}{3} r_h$, where $r_h$ is the 2-D, projected
half-light radius.

\begin{equation}
M_{1/2} = 4G^{-1} \sigma_v^2 r_h = 3G^{-1} \sigma_v^2 r_{1/2} \label{eq:mhalf}
\end{equation}

\noindent
We also calculated the mass-to-light ratio within $r_{1/2}$
($(M/L_V)_{1/2}$).  The errors on $M_{1/2}$ and $(M/L_V)_{1/2}$
include errors on the distance, $L_V$, $r_h$ and $\sigma_v$.

Fig.~\ref{fig:vhist} shows the velocity distributions of the seven
dwarf galaxies.  Non-member stars not included in the measurements of
the velocity dispersions are shaded.  The maximum likelihood values of
$\langle v_{\rmn{helio}} \rangle$ are shown as vertical dotted lines.
The solid curves are Gaussians with the maximum likelihood value of
$\sigma_v$.  The curves have been widened by the estimated
uncertainties on the radial velocities.  In detail, we constructed an
error kernel by stacking $N$ unit-area Gaussians, where $N$ is the
number of member stars.  Each Gaussian in the stack had a width equal
to the velocity error of one member star.  The Gaussians representing
the velocity distributions were convolved with the error kernel before
plotting in Fig.~\ref{fig:vhist}.  The bottom right panel of
Fig.~\ref{fig:vhist} shows the error kernel.

\subsection{Effect of membership}
\label{sec:sigmavtest}

\begin{figure}
 \centering
 \includegraphics[width=85mm]{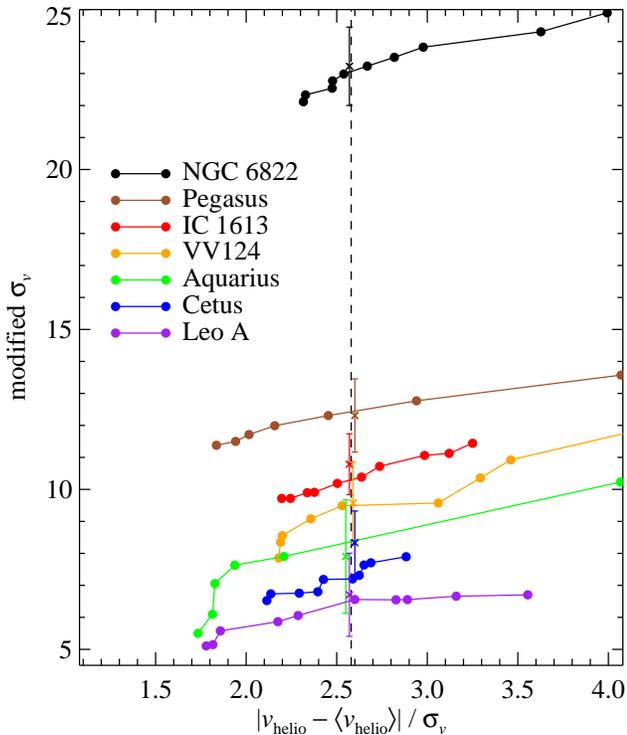}
 \caption{The effect on the velocity dispersion of excluding or
   including additional stars as members (see
   Sec.~\ref{sec:sigmavtest}).  The modified dispersions increase as
   additional stars are included.  The $x$-axis shows the absolute
   difference between the added star's velocity and the galaxy's mean
   velocity in units of the originally measured velocity dispersion,
   shown as crosses (horizontally shifted for clarity).  The dashed
   line shows our adopted membership cut, $2.58\sigma_v$.  The crosses
   do not always intersect the curves at the dashed line because our
   final membership cut allows stars to the right of the line if their
   measurement uncertainties encompass
   $2.58\sigma_v$.\label{fig:sigmavtest}}
\end{figure}

The measurement of velocity dispersion depends on the membership
criteria.  A strict membership cut generally leads to a lower
$\sigma_v$, whereas including stars at the fringes of the velocity
distribution can inflate $\sigma_v$.  Unfortunately, those same stars
are also the stars with the most uncertain membership.

We explored the effect on $\sigma_v$ of discarding and including stars
in the membership lists.  We started with a membership list with all
stars within $2.58\sigma_v$ of $\langle v_{\rmn{helio}} \rangle$.
This membership list is not identical to that described in
Sec.~\ref{sec:membership}, which included stars beyond $2.58\sigma_v$
as long as their error bars encompassed $\langle v_{\rmn{helio}}
\rangle$.  For simplicity, the membership cut for the purposes of this
test is a strict cut regardless of the stars' velocity uncertainties.
From this list, we removed the star with the velocity farthest from
$\langle v_{\rmn{helio}} \rangle$ and recomputed $\sigma_v$ with
$10^6$ MCMC trials.  We continued removing up to five stars.  Then we
added up to five stars to the strict membership list with velocities
more than $2.58\sigma_v$ discrepant from $\langle v_{\rmn{helio}}
\rangle$.

Fig.~\ref{fig:sigmavtest} shows the result of this test.  As expected,
$\sigma_v$ increases as stars farther removed from $\langle
v_{\rmn{helio}} \rangle$ are included.  The effect of adding an
additional star depends not only on its deviance from $\langle
v_{\rmn{helio}} \rangle$ but also its measurement uncertainty.  For
example, the extra stars in NGC~6822 have low velocity uncertainties.
Therefore, adding them steadily increases $\sigma_v$.  On the other
hand, the extra stars in Leo~A have large velocity uncertainties.
Adding them has only a small effect on $\sigma_v$.

Because our final membership cut is not a strict cut at
$2.58\sigma_v$, our final determinations of $\sigma_v$ (crosses in
Fig.~\ref{fig:sigmavtest}) do not always intersect the curves at
$2.58\sigma_v$.  The effect of our soft membership cut is most
apparent for Cetus, where our measurement of $\sigma_v$ is about one
standard deviation higher than if it were based on a strict membership
cut.  The soft membership cut is especially appropriate for galaxies,
like Cetus, with velocity dispersions on the same order as the
velocity uncertainties for individual stars.  A strict membership cut
for such galaxies would discard a larger fraction of stars than for
galaxies with comparatively large $\sigma_v$.  Even so, the difference
in $\sigma_v$ between a strict and soft membership cut is at most
about 1~km~s$^{-1}$.

Adding or removing stars between $2\sigma_v$ and $3.5\sigma_v$ affects
$\sigma_v$ by about $\pm 1$~km~s$^{-1}$, which is on the order of the
error on $\sigma_v$.  We conclude that the choice of membership is
important, but the exclusion of inclusion of a few stars does not
alter $\sigma_v$ by more than the errors quoted in
Table~\ref{tab:summary}.

\subsection{Rotation}
\label{sec:pegrot}

\begin{figure}
 \centering
 \includegraphics[width=85mm]{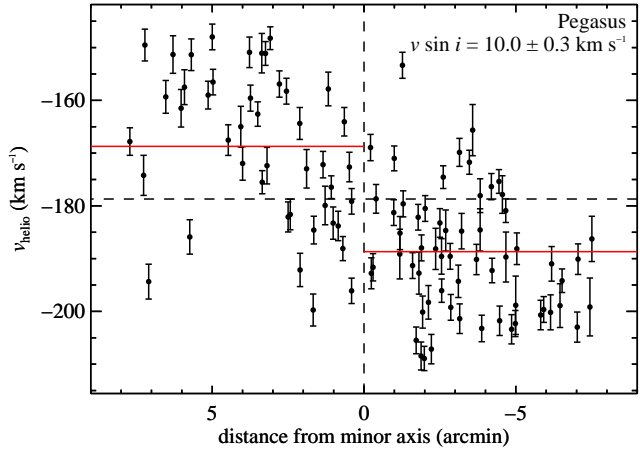}
 \caption{Rotation curve of Pegasus.  The solid red lines show the
   mean velocity on the east (left/positive) and west (right/negative)
   sides of the major axis.\label{fig:pegrot}}
\end{figure}

Fig.~\ref{fig:pegrot} shows the stellar velocities in Pegasus as a
function of displacement along the major axis, assuming a sky position
of $\alpha_0 = 23^{\rmn h} 28^{\rmn m} 36^{\rmn s}$, $\delta_0 =
+14\degr 44\arcmin 35\arcsec$ and a position angle of $122\degr$
\citep{hun06}.  The stars are clearly rotating.  We calculated the
mean $v_{\rmn{helio}}$ separately on the east and west sides of the
minor axis.  Half of the difference between the two velocities is
$\pegvrot \pm \pegvroterr$~km~s$^{-1}$, where the error is the
standard error on the mean.  We take this value to be the projected
rotation of the stars, $v \sin i$.

\citet{you03} measured the H$\,${\sc i} density and velocity
distribution of Pegasus.  They also found a velocity gradient very
similar to Fig.~\ref{fig:pegrot}.  They suggested that bubbles and
random motions -- rather than rotation -- cause the velocity gradient
in the gas because the gas is clumpy, and the density of gas on the
east side of the galaxy is larger than on the west side.  However, we
found that the velocity gradient is also present in red giants.  If
the gas gradient were caused by short-term hydrodynamical events, like
winds from a supernova, then those events would not affect the
kinematics of the red giants, which were presumably born before the
recent supernova.  Instead, the stars are moving in the same direction
and at the same velocity as the gas \citep[see fig.~6 of][]{you03}.
Therefore, we suggest that the H$\,${\sc i} gas is in fact rotating.

We modified our method of calculating $\sigma_v$ in Pegasus to account
for the rotation.  We subtracted $v \sin i$ from the velocities of the
stars in the eastern half of Pegasus, and we added $v \sin i$ to the
western stars.  In other words, we removed the rotation.  The velocity
histogram for Pegasus in Fig.~\ref{fig:vhist} reflects this
modification, which reduces $\sigma_v$.  We also modified the
calculation of $M_{1/2}$ to account for rotation support in addition
to pressure support.  We replaced $\sigma_v^2$ in Eq.~\ref{eq:mhalf}
with $\sigma_v^2 + \frac{1}{2}(v \sin i)^2$ \citep{wei06}.  The
coefficient on $(v \sin i)^2$ depends on the mass profile, but its
exact value matters less than uncertainty in inclination and the
assumption of spherical symmetry implicit in Eq.~\ref{eq:mhalf}.

The stellar velocity distribution of Pegasus, shown in
Fig.~\ref{fig:pegrot}, might be modelled as solid body rotation
(velocity linearly increasing with radius) just as well as flat
rotation.  Pegasus is part of Local Irregulars That Trace Luminosity
Extremes, The H$\,${\sc i} Nearby Galaxy Survey \citep[LITTLE
  THINGS,][]{hun12}, a detailed H$\,${\sc i} survey of dIrrs with
velocity resolution of 2.6~km~s$^{-1}$.  A prospect for future work is
to construct a dynamical model of Pegasus that combines our stellar
velocities with the LITTLE THINGS gas map.  That approach would allow
not only a more accurate measurement of the dynamical mass of Pegasus
but also a mass profile as function of radius.

We checked all of the other dwarf galaxies in our sample for stellar
rotation.  We found Pegasus to be the only galaxy in our sample with
obviously rotating stars.  (The gas may rotate or exhibit velocity
structure independently from the stars, as shown by \citeauthor{lo93}
\citeyear{lo93}.  However, the gas and stars are independent tracers
of mass.  Gas rotation does not affect our conclusions based on
stellar motions.)  \citet{dem06} found that the carbon stars in the
spheroid of NGC~6822 rotate perpendicular to the H$\,${\sc i} disc.
The rotation curve seems to increase with radius.  Their sample
spanned an area of the galaxy about 2.5~times larger than our DEIMOS
sample.  Our more centrally concentrated sample has a hint of some
velocity structure.  It is possible that rotation on the order of
$\sim 10$~km~s$^{-1}$ is present in our data, but the large velocity
dispersion obscures the signal of rotation.

Two of the isolated galaxies that we did not observe are known to
rotate.  \citet{lea12} measured a stellar rotation velocity of $17 \pm
1$~km~s$^{-1}$ in WLM.  They constructed a dynamical model of the
rotation of WLM in order to calculate $M_{1/2}$.  Footnote $b$ of
Table~\ref{tab:summary} summarises their results.  Their model is more
appropriate than our simple modification to \citeauthor{wol10}'s
(\citeyear{wol10}) formula to account for rotation.  \citet{fra09}
also measured rotation in Tucana.  Tucana is faint ($6 \times
10^5~\lsun$), which makes building a spectroscopic sample challenging.
As a result, the measurement of the rotation velocity is highly
uncertain.  Assuming a flat rotation curve, the rotation velocity is
about 15~km~s$^{-1}$.  We assume an uncertainty of 5~km~s$^{-1}$.

Four satellites of M31 are also known to rotate.  NGC~205
\citep{geh06} is too luminous for our consideration.  NGC~147 and
NGC~185 are dwarf elliptical galaxies with rotation velocities of $17
\pm 2$ and $15 \pm 5$~km~s$^{-1}$, respectively \citep{geh10}.
Finally, the dSph And~II exhibits prolate rotation with a velocity of
$10.9 \pm 2.4$~km~s$^{-1}$ \citep{ho12}.  As with Pegasus, we
incorporated the rotation velocities of WLM, Tucana, NGC~147, NGC~185,
and And~II into the derivation of $M_{1/2}$.

\subsection{Comparison to previous measurements}

Three of the galaxies in our sample have previous stellar velocity
dispersion measurements in the literature.  \citet{tol01} measured
$\sigma_v = 24.5$~km~s$^{-1}$ from 23~red giants in NGC~6822.  They
did not quote an uncertainty because they used the velocity dispersion
only to determine membership.  Regardless, their measurement is close
to our value ($\sigma_v = \nsettsigmav \pm
\nsettsigmaverr$~km~s$^{-1}$).

\citet{lew07} measured $\sigma_v = 17 \pm 2$~km~s$^{-1}$ from red
giants in Cetus.  This measurement is approximately
$\cetsigmavsigmadiff\sigma$ above our measurement of $\sigma_v =
\cetsigmav \pm \cetsigmaverru$~km~s$^{-1}$.  Part of the discrepancy
is due to the classification of member stars.  When we applied the
same membership criteria (Sec.~\ref{sec:membership}) and the same
technique to measure $\sigma_v$ (Sec.~\ref{sec:disp}) to
\citeauthor{lew07}'s catalogue of velocities, we obtained $\sigma_v =
\lewsigmav^{+\lewsigmaverru}_{-\lewsigmaverrl}$~km~s$^{-1}$, which
lessens the difference between the samples to
$\lewsigmavsigmadiff\sigma$.  The remaining difference may be due to
the higher S/N of our sample and to differences in the details of
measuring the velocities of individual stars.  See
Sec.~\ref{sec:lew07} and Fig.~\ref{fig:lew07} for a comparison of our
measurements of individual stellar velocities to those of
\citet{lew07}.

\citet{bro07} measured $\sigma_v = 9.3 \pm 1.3$~km~s$^{-1}$ from ten B
supergiants and two H$\,${\sc ii} regions in Leo~A\@.  They calculated
this dispersion by subtracting the average measurement uncertainty in
quadrature from the root mean square of the velocities.  Their
measurement is about $\leoasigmavsigmadiff\sigma$ above our
measurement of $\sigma_v =
\leoasigmav^{+\leoasigmaverru}_{-\leoasigmaverrl}$~km~s$^{-1}$.  We
applied an MCMC with maximum likelihood (Eq.~\ref{eq:l}) to
\citeauthor{bro07}'s data, and we determined $\sigma_v =
\brosigmav^{+\brosigmaverru}_{-\brosigmaverrl}$~km~s$^{-1}$, which is
still a difference of $\brosigmavsigmadiff\sigma$ from our measurement
of $\sigma_v$ from red giants.  Although the discrepancy is not highly
significant, it may be indicating an interesting difference between
the dynamics of the young (B supergiants and H$\,${\sc ii} regions)
and old or intermediate-aged (red giants) stellar populations in
Leo~A\@.  The apparent decrease of velocity dispersion with age is in
contrast to the observed increase of velocity dispersion with age in
WLM \citep{lea12}.


\section{Discussion}
\label{sec:discussion}

\begin{figure*}
 \centering
 \begin{minipage}{85mm}
 \centering
 \includegraphics[width=85mm]{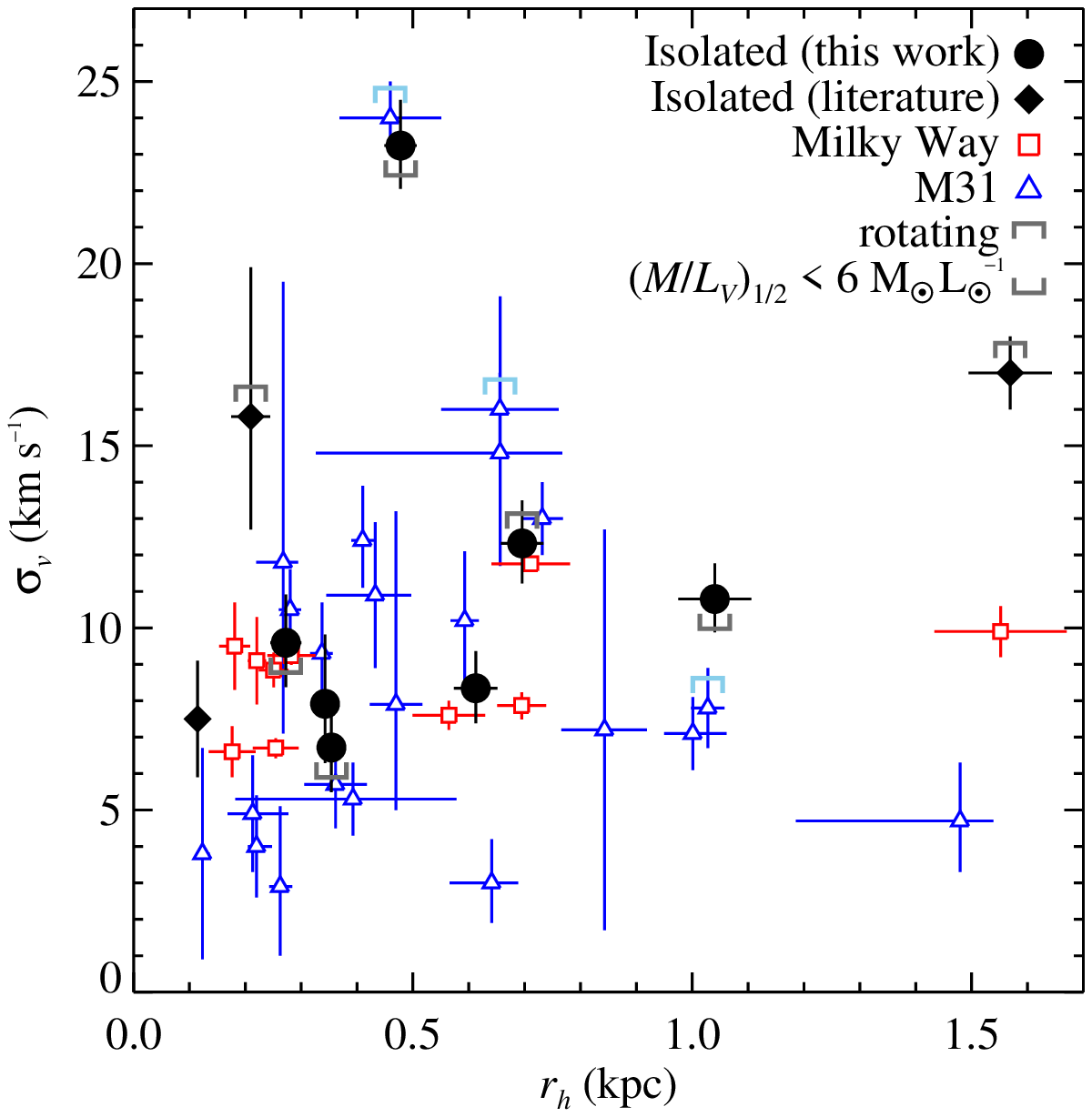}
 \end{minipage}
 \begin{minipage}{85mm}
 \centering
 \includegraphics[width=85mm]{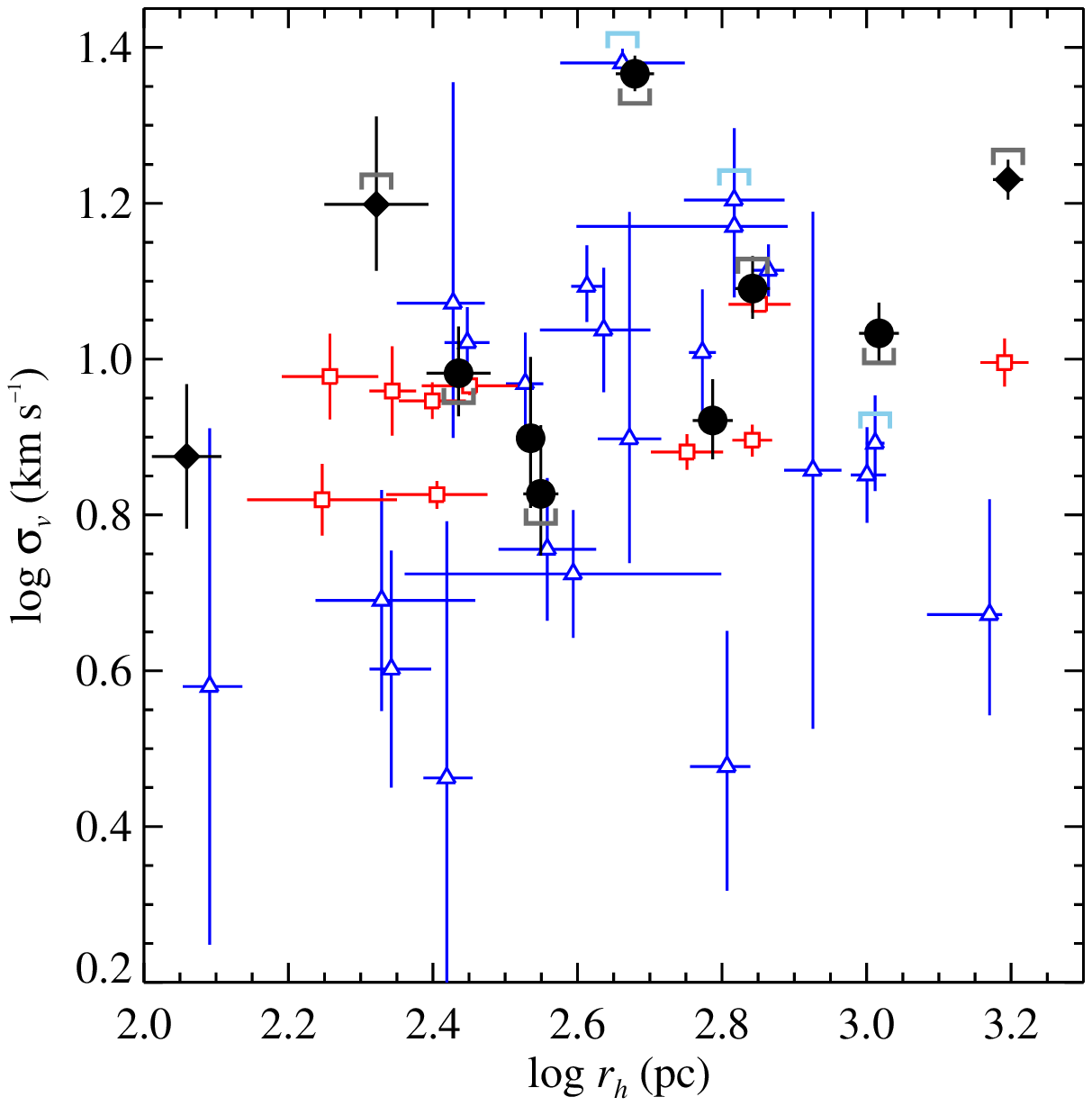}
 \end{minipage}
 \caption{Velocity dispersion versus projected half-light radius for
   dwarf galaxies in the field (black circles), satellites of the
   Milky Way (red squares) and satellites of M31 (blue triangles).
   Only galaxies with $10^5 < L_V/\lsun < 2 \times 10^8$ are shown.
   Both panels show the same data on linear (left) and logarithmic
   (right) axes.  The central masses of galaxies indicated with marks
   under the points have a significant component from baryons
   ($(M/L_V)_{1/2} < 6~\msun~\lsun^{-1}$).  The velocity dispersions
   for these galaxies from dark matter alone would be lower.  Rotating
   galaxies are indicated by marks over the points.  For these
   galaxies, the rotation-corrected velocities for mass estimation
   (see Sec.~\ref{sec:pegrot}) are 15--40 per cent larger than
   $\sigma_v$.\label{fig:sigmavrhalf}}
\end{figure*}

A comparison of the kinematic and structural properties of isolated
dwarf galaxies to those of the dwarf satellites of the Milky Way and
M31 has the potential to shed light on a number of issues related to
galaxy formation.  Perhaps the leading model for the formation of
dSphs is that these galaxies are the descendants of dIrrs that have
been tidally harassed and stripped of their gas as a result of falling
into a more massive halo \citep[e.g.,][]{lin83,may01,kor12}.  Our
results, as well as comparative studies of the metallicities of the
two populations \citep{kir13b}, place important constraints on this
and related models of dwarf galaxy transformations.  Similarly, they
inform models that invoke strong tidal stripping and mass loss to
explain the TBTF problem.

Fig.~\ref{fig:sigmavrhalf} shows $\sigma_v$ versus the
two-dimensional, projected $r_h$ for both the isolated galaxies and
satellites with $10^5 < L_V/\lsun < 2 \times 10^8$.  This luminosity
cut is intended to restrict the samples to about the same range of
stellar mass.  These data are the same as those presented in
Table~\ref{tab:summary}.  The $y$-axis in Fig.~\ref{fig:sigmavrhalf}
is intended to be a rough proxy for the mass of the galaxy.  Marks
over points indicate galaxies with significant rotation (Pegasus, WLM,
Tucana, and And~II).  A complete estimate of the mass would include
the rotation.  The values of $M_{1/2}$ in Table~\ref{tab:summary}
include rotation by replacing $\sigma_v^2$ in Eq.~\ref{eq:mhalf} with
$\sigma_v^2 + \frac{1}{2}(v \sin i)^2$ \citep[][see
  Sec.~\ref{sec:pegrot}]{wei06}.  The modified velocity dispersion is
15--40 per cent larger than $\sigma_v$ for the rotating galaxies.

We use stellar velocities to trace the mass distribution.  For many
galaxies, the mass distribution is dominated by dark matter.  However,
the stars of some galaxies are centrally concentrated enough that the
stellar velocities are about as sensitive to stellar mass as dark
matter mass.  Furthermore, most of the isolated dIrrs have gas masses
of the same order as the stellar masses.  Mass models of the dark
matter subhaloes must take these baryons into account.
Fig.~\ref{fig:sigmavrhalf} identifies galaxies where baryons are
especially important ($(M/L_V)_{1/2} < 6~\msun~\lsun^{-1}$).

The two samples are not distinct in the $\sigma_v$--$r_h$ plane.  A
two-dimensional Kolmogorov--Smirnov (K--S) test yields a
$\ksallrh$~per cent probability that the isolated galaxies are drawn
from the same parent population as Milky Way and M31 satellites.  This
statistic increases to $\ksallrhmod$~per cent if $\sqrt{\sigma_v^2 +
  \frac{1}{2}(v \sin i)^2}$ is used for the four rotating galaxies
instead of $\sigma_v$.  However, we caution that empirical
distribution function tests, like the K--S and Anderson--Darling
tests, are technically invalid in more than one dimension because
there is not a unique ordering of data points \citep{fei12}.

Even in the absence of a rigorous statistical test,
Fig.~\ref{fig:sigmavrhalf} shows that there is no obvious distinction
between the isolated dwarf galaxies and the satellites.  WLM is the
farthest outlying galaxy, with both a high $\sigma_v$ and a high $r_h$
compared to the bulk of the other galaxies.  At least part of the
difference is due to the fact that WLM is a dIrr whereas all of the
satellite galaxies in Table~\ref{tab:summary} and
Fig.~\ref{fig:sigmavrhalf} are dSphs.  Besides WLM, other dIrrs have
some distinctions from the dSphs.  Although there may be hints of
rotation in Fornax and Sculptor \citep{wal06a,bat08}, only some of the
most luminous M31 dSphs -- NGC~147, NGC~185, NGC~205, and And~II
\citep{geh06,geh10,ho12} -- have clear stellar rotation curves.  On
the other hand, dIrrs as faint as Pegasus ($L_V = 7 \times
10^6~\lsun$) and Tucana ($L_V = 6 \times 10^5~\lsun$) are rotating.
Whereas dSphs mostly have flat velocity dispersions as a function of
radius \citep[e.g.,][]{wal06a,wal06b,bat08}, the velocity dispersion
of WLM decreases with radius \citep{lea12}.  The ellipticities of the
dIrrs are larger on average than the dSphs, perhaps due to the
presence of rotating discs in some dIrrs.  The most obvious difference
is the presence of gas in the dIrrs but not the dSphs \citep{grc09}.
Together, all of these pieces of evidence will provide strong
constraints on models of the formation and evolution of dwarf
galaxies, especially the possible transformation of dIrrs to dSphs.

Although proximity to a large host definitely influences the
kinematics (support by rotation versus dispersion), structure
(ellipticity) and gas content of dwarf galaxies, it is not clear that
environment can explain the TBTF problem.  TBTF can be viewed in terms
of the maximum circular velocity of a subhalo ($v_{\rmn{max}}$) and
its radius when it achieved that circular velocity ($r_{\rmn{max}}$).
Both $v_{\rmn{max}}$ and $r_{\rmn{max}}$ are derived from the directly
observable quantities $\sigma_v$ and $r_h$.  Because the isolated and
satellite galaxies are not obviously distinct in the $\sigma_v$--$r_h$
plane, environment is not an obvious cause of TBTF\@.

Another way to frame TBTF is that dark matter simulations predict more
dense satellites than are observed.  However, the field of the Local
Group has no galaxy denser than the densest satellite of the MW or
M31.  Therefore, the isolated galaxies, which are minimally affected
by the gravitational and ram pressure influences of the large spiral
galaxies, also exhibit the same range of structural properties that
give rise to the TBTF problem for satellite galaxies.

\begin{figure}
 \centering
 \includegraphics[width=85mm]{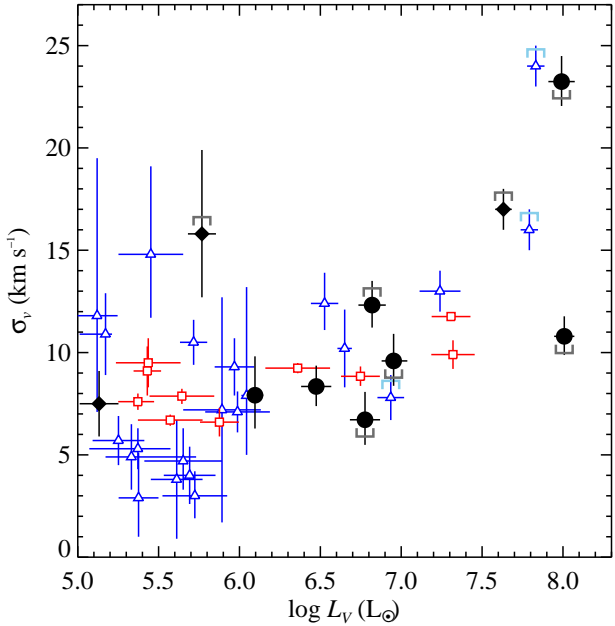}
 \caption{Velocity dispersion versus luminosity for Local Group dwarf
   galaxies.  The symbols are the same as in
   Fig.~\ref{fig:sigmavrhalf}.\label{fig:sigmavlum}}
\end{figure}

Of course, the Milky Way and M31 do tidally disturb some of their dSph
satellites, like Sagittarius \citep*{iba94} and Hercules
\citep{dea12}.  \citet{brozol12} predicted that these tidal forces
would cause satellite galaxies to have a lower circular velocity than
field dwarf galaxies in the same luminosity range.
Fig.~\ref{fig:sigmavlum} shows velocity dispersions (not circular
velocities) versus luminosities for both field and satellite dwarf
galaxies.  Circular velocities and velocity dispersions are not
proportional when the stars are rotating, but most of the galaxies in
our sample do not have stellar rotation.  The galaxies do not separate
any more in this space than in the space of velocity dispersion versus
half-light radius.  The 2-D K--S test between the field and satellite
galaxies with $L_V/\lsun > 10^6$ returns a probability of $\ksalllum$
per cent that the galaxies are drawn from the same population.
Accounting for rotational support reduces the probability only to
$\ksalllummod$ per cent.  Therefore, our observations impose
limitations on both (1) proposed mechanisms for the transformation of
dIrrs into dSphs and (2) environmental solutions to TBTF whether the
problem is considered in terms of half-light radius or luminosity.

We have considered only one dynamical tracer population: red giants.
All of the galaxies in our sample except Cetus also have gas.  We have
also made only the most basic estimate of dynamical mass ($M_{1/2}$).
A worthwhile prospect for future work is to construct detailed models
of the mass profiles of the galaxies we observed.  For example,
\citet{ada12} constructed such a model for NGC~2976.  Our individual
stellar velocities are available in Table~\ref{tab:vel} for interested
modellers.  Gas maps from LITTLE THINGS \citep{hun12} add value to the
stellar velocities.  Taken together, the stellar and gas kinematics
can be used to make some of the most detailed mass profiles of
galaxies yet.  These profiles would be relevant to understanding the
role of environment in the formation and evolution of dwarf galaxies,
solving the TBTF problem and determining whether dwarf galaxies have
cusped or cored dark matter profiles.

\section*{Acknowledgements}

We are grateful to the many people who have worked to make the Keck
Telescopes and their instruments a reality and to operate and maintain
the Keck Observatory.  The authors wish to extend special thanks to
those of Hawaiian ancestry on whose sacred mountain we are privileged
to be guests.  Without their generous hospitality, none of the
observations presented herein would have been possible.

We thank Josh Simon and Marla Geha for providing their DEIMOS radial
velocity template spectra.  We also thank Alan McConnachie and Mike
Irwin for sharing their photometric and astrometric catalogue for
Cetus.  ENK acknowledges support from the Southern California Center
for Galaxy Evolution, a multicampus research program funded by the
University of California Office of Research, and partial support from
NSF grant AST-1009973.  JGC thanks NSF grant AST-0908139 for partial
support.

\label{lastpage}

\begin{thebibliography}{99}

\bibitem[\protect\citeauthoryear{Adams et al.}{2012}]{ada12} Adams
  J.~J., Gebhardt K., Blanc G.~A., Fabricius M.~H., Hill G.~J., Murphy
  J.~D., van den Bosch R.~C.~E., van de Ven G., 2012, ApJ, 745, 92

\bibitem[\protect\citeauthoryear{Amorisco, Zavala \& de Boer}{Amorisco
    et al.}{2013}]{amo13} Amorisco N.~C., Zavala J., de Boer T.~J.~L.,
  2013, ApJL, submitted (arXiv:1309.5958)

\bibitem[\protect\citeauthoryear{Appenzeller et al.}{1998}]{app98}
  Appenzeller I., et al., 1998, Messenger, 94, 1

\bibitem[\protect\citeauthoryear{Arraki et al.}{2013}]{arr13} Arraki
  K.~S., Klypin A., More S., Trujillo-Gomez S., 2013, MNRAS, 2956

\bibitem[\protect\citeauthoryear{Battaglia et al.}{2008}]{bat08}
  Battaglia G., Helmi A., Tolstoy E., Irwin M., Hill V., Jablonka P.,
  2008, ApJ, 681, L13

\bibitem[\protect\citeauthoryear{Bell, Slater, \&
    Martin}{2011}]{bell11} Bell E.~F., Slater C.~T., Martin N.~F.,
  2011, ApJ, 742, L15

\bibitem[\protect\citeauthoryear{Bellazzini et al.}{2004}]{bel04}
  Bellazzini M., Gennari N., Ferraro F.~R., Sollima A., 2004, MNRAS,
  354, 708

\bibitem[\protect\citeauthoryear{Bellazzini, Gennari, \&
    Ferraro}{2005}]{bel05} Bellazzini M., Gennari N., Ferraro F.~R.,
  2005, MNRAS, 360, 185

\bibitem[\protect\citeauthoryear{Bellazzini et al.}{2011}]{bel11}
  Bellazzini M., et al., 2011, A\&A, 527, A58

\bibitem[\protect\citeauthoryear{Bernard et al.}{2009}]{ber09} Bernard
  E.~J., et al., 2009, ApJ, 699, 1742

\bibitem[\protect\citeauthoryear{Bernard et al.}{2010}]{ber10} Bernard
  E.~J., et al., 2010, ApJ, 712, 1259

\bibitem[\protect\citeauthoryear{Bernard et al.}{2013}]{ber13} Bernard
  E.~J., et al., 2013, MNRAS, 432, 3047

\bibitem[\protect\citeauthoryear{Binggeli, Tarenghi, \&
    Sandage}{Binggeli et al.}{1990}]{bin90} Binggeli B., Tarenghi M.,
  Sandage A., 1990, A\&A, 228, 42

\bibitem[\protect\citeauthoryear{Bonanos et al.}{2004}]{bon04} Bonanos
  A.~Z., Stanek K.~Z., Szentgyorgyi A.~H., Sasselov D.~D., Bakos
  G.~{\'A}., 2004, AJ, 127, 861

\bibitem[\protect\citeauthoryear{Boylan-Kolchin, Bullock \&
    Kaplinghat}{Boylan-Kolchin et al.}{2011}]{boy11} Boylan-Kolchin
  M., Bullock J.~S., Kaplinghat M., 2011, MNRAS, 415, L40

\bibitem[\protect\citeauthoryear{Boylan-Kolchin, Bullock \&
    Kaplinghat}{Boylan-Kolchin et al.}{2012}]{boy12} Boylan-Kolchin
  M., Bullock J.~S., Kaplinghat M., 2012, MNRAS, 422, 1203

\bibitem[\protect\citeauthoryear{Brasseur et al.}{2011}]{bra11}
  Brasseur C.~M., Martin N.~F., Rix H.-W., Irwin M., Ferguson
  A.~M.~N., McConnachie A.~W., de Jong J., 2011, ApJ, 729, 23

\bibitem[\protect\citeauthoryear{Brooks et al.}{2013}]{bro13} Brooks
  A.~M., Kuhlen M., Zolotov A., Hooper D., 2013, ApJ, 765, 22

\bibitem[\protect\citeauthoryear{Brooks \& Zolotov}{2012}]{brozol12}
  Brooks A.~M., Zolotov A., 2012, ApJ, submitted (arXiv:1207.2468)

\bibitem[\protect\citeauthoryear{Brown et al.}{2007}]{bro07} Brown
  W.~R., Geller M.~J., Kenyon S.~J., Kurtz M.~J., 2007, ApJ, 666, 231

\bibitem[\protect\citeauthoryear{Carrera et al.}{2002}]{car02} Carrera
  R., Aparicio A., Mart{\'{\i}}nez-Delgado D., Alonso-Garc{\'{\i}}a
  J., 2002, AJ, 123, 3199

\bibitem[\protect\citeauthoryear{Cohen}{1978}]{coh78} Cohen J.~G.,
  1978, ApJ, 221, 788

\bibitem[\protect\citeauthoryear{Collins et al.}{2010}]{col10} Collins
  M.~L.~M., et al., 2010, MNRAS, 407, 2411

\bibitem[\protect\citeauthoryear{Collins et al.}{2013}]{col13} Collins
  M.~L.~M., et al., 2013, ApJ, 768, 172

\bibitem[\protect\citeauthoryear{Conn et al.}{2012}]{con12} Conn
  A.~R., et al., 2012, ApJ, 758, 11

\bibitem[\protect\citeauthoryear{Cook et al.}{1999}]{coo99} Cook
  K.~H., Mateo M., Olszewski E.~W., Vogt S.~S., Stubbs C., Diercks A.,
  1999, PASP, 111, 306

\bibitem[\protect\citeauthoryear{Cooper et al.}{2012}]{coo12} Cooper
  M.~C., Newman J.~A., Davis M., Finkbeiner D.~P., Gerke B.~F., 2012,
  Astrophysics Source Code Library, record ascl:1203.003, 3003

\bibitem[\protect\citeauthoryear{Dalcanton et al.}{2012}]{dal12}
  Dalcanton J.~J., et al., 2012, ApJS, 200, 18

\bibitem[\protect\citeauthoryear{Dale et al.}{2007}]{dal07} Dale
  D.~A., et al., 2007, ApJ, 655, 863

\bibitem[\protect\citeauthoryear{de Jong et al.}{2008}]{dej08} de Jong
  J.~T.~A., et al., 2008, ApJ, 680, 1112

\bibitem[\protect\citeauthoryear{de Vaucouleurs et al.}{1991}]{dev91}
  de Vaucouleurs G., de Vaucouleurs A., Corwin H.~G., Jr., Buta R.~J.,
  Paturel G., Fouqu{\'e} P., 1991, Third Reference Catalogue of Bright
  Galaxies, Vol.\ 1--3. Springer, Berlin

\bibitem[\protect\citeauthoryear{Deason et al.}{2012}]{dea12} Deason
  A.~J., Belokurov V., Evans N.~W., Watkins L.~L., Fellhauer M., 2012,
  MNRAS, 425, L101

\bibitem[\protect\citeauthoryear{Demers, Battinelli \& Kunkel}{Demers
    et al.}{2006}]{dem06} Demers S., Battinelli P., Kunkel W.~E.,
  2006, ApJ, 636, L85


\bibitem[\protect\citeauthoryear{Faber et al.}{2003}]{fab03} Faber
  S.~M., et al., 2003, Proc.\ SPIE, 4841, 1657

\bibitem[\protect\citeauthoryear{Feigelson \& Babu}{2012}]{fei12}
  Feigelson E.~D., Babu J.~G., 2012, Modern Statistical Methods for
  Astronomy. Cambridge Univ.\ Press, Cambridge

\bibitem[\protect\citeauthoryear{Fraternali et al.}{2009}]{fra09}
  Fraternali F., Tolstoy E., Irwin M.~J., Cole A.~A., 2009, A\&A, 499,
  121

\bibitem[\protect\citeauthoryear{Frinchaboy et al.}{2012}]{fri12}
  Frinchaboy P.~M., Majewski S.~R., Mu{\~n}oz R.~R., Law D.~R.,
  {\L}okas E.~L., Kunkel W.~E., Patterson R.~J., Johnston K.~V., 2012,
  ApJ, 756, 74

\bibitem[\protect\citeauthoryear{Garrison-Kimmel et al.}{2013}]{gar13}
  Garrison-Kimmel S., Rocha M., Boylan-Kolchin M., Bullock J.~S.,
  Lally J., 2013, MNRAS, 433, 3539

\bibitem[\protect\citeauthoryear{Geha et al.}{2006}]{geh06} Geha M.,
  Guhathakurta P., Rich R.~M., Cooper M.~C., 2006, AJ, 131, 332

\bibitem[\protect\citeauthoryear{Geha et al.}{2010}]{geh10} Geha M.,
  van der Marel R.~P., Guhathakurta P., Gilbert K.~M., Kalirai J.,
  Kirby E.~N., 2010, ApJ, 711, 361

\bibitem[\protect\citeauthoryear{Gieren et al.}{2006}]{gie06} Gieren
  W., Pietrzy{\'n}ski G., Nalewajko K., Soszy{\'n}ski I., Bresolin F.,
  Kudritzki R.-P., Minniti D., Romanowsky A., 2006, ApJ, 647, 1056

\bibitem[\protect\citeauthoryear{Gilbert et al.}{2006}]{gil06} Gilbert
  K.~M., et al., 2006, ApJ, 652, 1188

\bibitem[\protect\citeauthoryear{Girardi et al.}{2002}]{gir02} Girardi
  L., Bertelli G., Bressan A., Chiosi C., Groenewegen M.~A.~T., Marigo
  P., Salasnich B., Weiss A., 2002, A\&A, 391, 195

\bibitem[\protect\citeauthoryear{Grcevich \& Putman}{2009}]{grc09}
  Grcevich J., Putman M.~E., 2009, ApJ, 696, 385

\bibitem[\protect\citeauthoryear{Grebel, Gallagher, \& Harbeck}{Grebel
    et al.}{2003}]{gre03} Grebel E.~K., Gallagher J.~S., III, Harbeck
  D., 2003, AJ, 125, 1926

\bibitem[\protect\citeauthoryear{Guhathakurta et al.}{2005}]{guh05}
  Guhathakurta P., Ostheimer J.~C., Gilbert K.~M., Rich R.~M.,
  Majewski S.~R., Kalirai J.~S., Reitzel D.~B., Patterson R.~J., 2005,
  preprint (arXiv:astro-ph/0502366)

\bibitem[\protect\citeauthoryear{Guhathakurta et al.}{2006}]{guh06}
  Guhathakurta P., et al., 2006, AJ, 131, 2497

\bibitem[\protect\citeauthoryear{Helmi et al.}{2012}]{hel12} Helmi A.,
  Sales L.~V., Starkenburg E., Starkenburg T.~K., Vera-Ciro C.~A., De
  Lucia G., Li Y.-S., 2012, ApJ, 758, L5

\bibitem[\protect\citeauthoryear{Hook et al.}{2004}]{hoo04} Hook
  I.~M., J{\o}rgensen I., Allington-Smith J.~R., Davies R.~L.,
  Metcalfe N., Murowinski R.~G., Crampton D., 2004, PASP, 116, 425

\bibitem[\protect\citeauthoryear{Hunter \& Elmegreen}{2006}]{hun06}
  Hunter D.~A., Elmegreen B.~G., 2006, ApJS, 162, 49

\bibitem[\protect\citeauthoryear{Hunter et al.}{2012}]{hun12} Hunter
  D.~A., et al., 2012, AJ, 144, 134

\bibitem[\protect\citeauthoryear{Ho et al.}{2012}]{ho12} Ho N., et
  al., 2012, ApJ, 758, 124

\bibitem[\protect\citeauthoryear{Ibata, Gilmore, \& Irwin}{Ibata et
    al.}{1994}]{iba94} Ibata R.~A., Gilmore G., Irwin M.~J., 1994,
  Nature, 370, 194

\bibitem[\protect\citeauthoryear{Ibata et al.}{2007}]{iba07} Ibata R.,
  Martin N.~F., Irwin M., Chapman S., Ferguson A.~M.~N., Lewis G.~F.,
  McConnachie A.~W., 2007, ApJ, 671, 1591

\bibitem[\protect\citeauthoryear{Irwin \&
    Hatzidimitriou}{1995}]{irw95} Irwin M., Hatzidimitriou D., 1995,
  MNRAS, 277, 1354

\bibitem[\protect\citeauthoryear{Irwin et al.}{2007}]{irw07} Irwin
  M.~J., et al., 2007, ApJ, 656, L13


\bibitem[\protect\citeauthoryear{Jacobs et al.}{2011}]{jac11} Jacobs
  B.~A., Tully R.~B., Rizzi L., Karachentsev I.~D., Chiboucas K., Held
  E.~V., 2011, EAS, 48, 67


\bibitem[\protect\citeauthoryear{Kirby, Guhathakurta \& Sneden}{Kirby
    et al.}{2008}]{kir08} Kirby E.~N., Guhathakurta P., Sneden C.,
  2008, ApJ, 682, 1217

\bibitem[\protect\citeauthoryear{Kirby et al.}{2009}]{kir09} Kirby
  E.~N., Guhathakurta P., Bolte M., Sneden C., Geha M.~C., 2009, ApJ,
  705, 328

\bibitem[\protect\citeauthoryear{Kirby, Cohen \& Bellazzini}{Kirby et
    al.}{2012}]{kir12} Kirby E.~N., Cohen J.~G., Bellazzini M., 2012,
  ApJ, 751, 46

\bibitem[\protect\citeauthoryear{Kirby et al.}{2013a}]{kir13a} Kirby
  E.~N., Boylan-Kolchin M., Cohen J.~G., Geha M., Bullock J.~S.,
  Kaplinghat M., 2013a, ApJ, 770, 16

\bibitem[\protect\citeauthoryear{Kirby et al.}{2013b}]{kir13b} Kirby
  E.~N., Cohen J.~G., Guhathakurta P., Cheng L., Bullock J.~S.,
  Gallazzi A., 2013b, ApJ, 779, 102

\bibitem[\protect\citeauthoryear{Knapp, Kerr, \& Bowers}{Knapp et
    al.}{1978}]{kna78} Knapp G.~R., Kerr F.~J., Bowers P.~F., 1978,
  AJ, 83, 360

\bibitem[\protect\citeauthoryear{Kormendy \& Bender}{2012}]{kor12}
  Kormendy J., Bender R., 2012, ApJS, 198, 2

\bibitem[\protect\citeauthoryear{Leaman et al.}{2009}]{lea09} Leaman
  R., Cole A.~A., Venn K.~A., Tolstoy E., Irwin M.~J., Szeifert T.,
  Skillman E.~D., McConnachie A.~W., 2009, ApJ, 699, 1

\bibitem[\protect\citeauthoryear{Leaman et al.}{2012}]{lea12} Leaman
  R., et al., 2012, ApJ, 750, 33

\bibitem[\protect\citeauthoryear{Lee}{1995}]{lee95} Lee M.~G., 1995,
  AJ, 110, 1129

\bibitem[\protect\citeauthoryear{Lee et al.}{2009}]{lee09} Lee M.~G.,
  Yuk I.-S., Park H.~S., Harris J., Zaritsky D., 2009, ApJ, 703, 692

\bibitem[\protect\citeauthoryear{Letarte et al.}{2010}]{let10} Letarte
  B., et al., 2010, A\&A, 523, A17

\bibitem[\protect\citeauthoryear{Lewis et al.}{2007}]{lew07} Lewis
  G.~F., Ibata R.~A., Chapman S.~C., McConnachie A., Irwin M.~J.,
  Tolstoy E., Tanvir N.~R., 2007, MNRAS, 375, 1364

\bibitem[\protect\citeauthoryear{Lin \& Faber}{1983}]{lin83} Lin
  D.~N.~C., Faber S.~M., 1983, ApJ, 266, L21

\bibitem[\protect\citeauthoryear{Lisker et al.}{2007}]{lis07} Lisker
  T., Grebel E.~K., Binggeli B., Glatt K., 2007, ApJ, 660, 1186

\bibitem[\protect\citeauthoryear{Lo, Sargent, \& Young}{1993}]{lo93}
  Lo K.~Y., Sargent W.~L.~W., Young K., 1993, AJ, 106, 507

\bibitem[\protect\citeauthoryear{Majewski et al.}{2003}]{maj03}
  Majewski S.~R., Skrutskie M.~F., Weinberg M.~D., Ostheimer J.~C.,
  2003, ApJ, 599, 1082

\bibitem[\protect\citeauthoryear{Majewski et al.}{2007}]{maj07}
  Majewski S.~R., et al., 2007, ApJ, 670, L9

\bibitem[\protect\citeauthoryear{Martin et al.}{2008}]{maretal08}
  Martin N.~F., et al., 2008, ApJ, 672, L13

\bibitem[\protect\citeauthoryear{Martin, de Jong, \&
    Rix}{2008}]{mar08} Martin N.~F., de Jong J.~T.~A., Rix H.-W.,
  2008, ApJ, 684, 1075

\bibitem[\protect\citeauthoryear{Martin et al.}{2009}]{mar09} Martin
  N.~F., et al., 2009, ApJ, 705, 758

\bibitem[\protect\citeauthoryear{Mateo, Olszewski, \&
    Walker}{2008}]{mat08} Mateo M., Olszewski E.~W., Walker M.~G.,
  2008, ApJ, 675, 201

\bibitem[\protect\citeauthoryear{Mayer et al.}{2001}]{may01} Mayer L.,
  Governato F., Colpi M., Moore B., Quinn T., Wadsley J., Stadel J.,
  Lake G., 2001, ApJ, 547, L123

\bibitem[\protect\citeauthoryear{McConnachie}{2012}]{mcc12}
  McConnachie A.~W., 2012, AJ, 144, 4

\bibitem[\protect\citeauthoryear{McConnachie et al.}{2006}]{mccetal06}
  McConnachie A.~W., Arimoto N., Irwin M., Tolstoy E., 2006, MNRAS,
  373, 715

\bibitem[\protect\citeauthoryear{McConnachie \& Irwin}{2006}]{mcc06}
  McConnachie A.~W., Irwin M.~J., 2006, MNRAS, 365, 1263

\bibitem[\protect\citeauthoryear{McConnachie et al.}{2005}]{mcc05}
  McConnachie A.~W., Irwin M.~J., Ferguson A.~M.~N., Ibata R.~A.,
  Lewis G.~F., Tanvir N., 2005, MNRAS, 356, 979

\bibitem[\protect\citeauthoryear{McConnachie et al.}{2008}]{mcc08}
  McConnachie A.~W., et al., 2008, ApJ, 688, 1009

\bibitem[\protect\citeauthoryear{McConnachie et al.}{2009}]{mcc09}
  McConnachie A.~W., et al., 2009, Nature, 461, 66

\bibitem[\protect\citeauthoryear{Monaco et al.}{2004}]{mon04} Monaco
  L., Bellazzini M., Ferraro F.~R., Pancino E., 2004, MNRAS, 353, 874

\bibitem[\protect\citeauthoryear{Newman et al.}{2013}]{new13} Newman
  J.~A., et al., 2013, ApJS, 208, 5

\bibitem[\protect\citeauthoryear{Pasquini et al.}{2002}]{pas02}
  Pasquini L., et al., 2002, Messenger, 110, 1

\bibitem[\protect\citeauthoryear{Pe{\~n}arrubia, Navarro \&
    McConnachie}{Pe{\~n}arrubia et al.}{2008}]{pen08} Pe{\~n}arrubia
  J., Navarro J.~F., McConnachie A.~W., 2008, ApJ, 673, 226

\bibitem[\protect\citeauthoryear{Pe{\~n}arrubia et al.}{2012}]{pen12}
  Pe{\~n}arrubia J., Pontzen A., Walker M.~G., Koposov S.~E., 2012,
  ApJ, 759, L42

\bibitem[\protect\citeauthoryear{Pietrzy{\'n}ski et al.}{2008}]{pie08}
  Pietrzy{\'n}ski G., et al., 2008, AJ, 135, 1993

\bibitem[\protect\citeauthoryear{Pietrzy{\'n}ski et al.}{2009}]{pie09}
  Pietrzy{\'n}ski G., G{\'o}rski M., Gieren W., Ivanov V.~D., Bresolin
  F., Kudritzki R.-P., 2009, AJ, 138, 459

\bibitem[\protect\citeauthoryear{Richardson et al.}{2011}]{ric11}
  Richardson J.~C., et al., 2011, ApJ, 732, 76

\bibitem[\protect\citeauthoryear{Sales et al.}{2007}]{sal07} Sales
  L.~V., Navarro J.~F., Abadi M.~G., Steinmetz M., 2007, MNRAS, 379,
  1475

\bibitem[\protect\citeauthoryear{Saviane, Held, \&
    Piotto}{1996}]{sav96} Saviane I., Held E.~V., Piotto G., 1996,
  A\&A, 315, 40

\bibitem[\protect\citeauthoryear{Schiavon et al.}{1997}]{sch97}
  Schiavon R.~P., Barbuy B., Rossi S.~C.~F., Milone A., 1997, ApJ,
  479, 902

\bibitem[\protect\citeauthoryear{Schlegel, Finkbeiner \&
    Davis}{Schlegel et al.}{1998}]{sch98} Schlegel D.~J., Finkbeiner
  D.~P., Davis M., 1998, ApJ, 500, 525

\bibitem[\protect\citeauthoryear{Simon \& Geha}{2007}]{sim07} Simon
  J.~D., Geha M., 2007, ApJ, 670, 313

\bibitem[\protect\citeauthoryear{Slater, Bell, \&
    Martin}{2011}]{sla11} Slater C.~T., Bell E.~F., Martin N.~F.,
  2011, ApJ, 742, L14

\bibitem[\protect\citeauthoryear{Spinrad \& Taylor}{1971}]{spi71}
  Spinrad H., Taylor B.~J., 1971, ApJS, 22, 445

\bibitem[\protect\citeauthoryear{Teyssier, Johnston \&
    Kuhlen}{Teyssier et al.}{2012}]{tey12} Teyssier M., Johnston
  K.~V., Kuhlen M., 2012, MNRAS, 426, 1808

\bibitem[\protect\citeauthoryear{Tollerud et al.}{2012}]{tol12}
  Tollerud E.~J., et al., 2012, ApJ, 752, 45

\bibitem[\protect\citeauthoryear{Tollerud et al.}{2013}]{tol13}
  Tollerud E.~J., Geha M.~C., Vargas L.~C., Bullock J.~S., 2013, ApJ,
  768, 50

\bibitem[\protect\citeauthoryear{Tolstoy et al.}{2001}]{tol01} Tolstoy
  E., Irwin M.~J., Cole A.~A., Pasquini L., Gilmozzi R., Gallagher
  J.~S., 2001, MNRAS, 327, 918

\bibitem[\protect\citeauthoryear{Walker et al.}{2006a}]{wal06a} Walker
  M.~G., Mateo M., Olszewski E.~W., Bernstein R., Wang X., Woodroofe
  M., 2006a, AJ, 131, 2114

\bibitem[\protect\citeauthoryear{Walker et al.}{2006b}]{wal06b} Walker
  M.~G., Mateo M., Olszewski E.~W., Pal J.~K., Sen B., Woodroofe M.,
  2006b, ApJ, 642, L41

\bibitem[\protect\citeauthoryear{Walker et al.}{2007}]{wal07} Walker
  M.~G., Mateo M., Olszewski E.~W., Gnedin O.~Y., Wang X., Sen B.,
  Woodroofe M., 2007, ApJ, 667, L53

\bibitem[\protect\citeauthoryear{Walker, Mateo, \&
    Olszewski}{2009a}]{wal09} Walker M.~G., Mateo M., Olszewski E.~W.,
  2009a, AJ, 137, 3100

\bibitem[\protect\citeauthoryear{Walker et al.}{2009b}]{waletal09}
  Walker M.~G., Mateo M., Olszewski E.~W., Pe{\~n}arrubia J., Wyn
  Evans N., Gilmore G., 2009b, ApJ, 704, 1274

\bibitem[\protect\citeauthoryear{Weiner et al.}{2006}]{wei06} Weiner
  B.~J., et al., 2006, ApJ, 653, 1027

\bibitem[\protect\citeauthoryear{Weisz et al.}{2011}]{wei11} Weisz
  D.~R., et al., 2011, ApJ, 739, 5

\bibitem[\protect\citeauthoryear{Wolf et al.}{2010}]{wol10} Wolf J.,
  Martinez G.~D., Bullock J.~S., Kaplinghat M., Geha M., Mu{\~n}oz
  R.~R., Simon J.~D., Avedo F.~F., 2010, MNRAS, 406, 1220

\bibitem[\protect\citeauthoryear{Young et al.}{2003}]{you03} Young
  L.~M., van Zee L., Lo K.~Y., Dohm-Palmer R.~C., Beierle M.~E., 2003,
  ApJ, 592, 111

\bibitem[\protect\citeauthoryear{Zolotov et al.}{2012}]{zol12} Zolotov
  A., et al., 2012, ApJ, 761, 71

\end{thebibliography}
\end{document}